\documentclass[iop]{emulateapj}
\usepackage{natbib}
\def\5{\footnotesize V\normalsize}
\def\4{\footnotesize IV\normalsize}
\def\3{\footnotesize III\normalsize}
\def\2{\footnotesize II\normalsize}
\def\1{\footnotesize I\normalsize}
\def\lam{$\lambda$}
\def\kms{$\mbox{km s}^{-1}$}

\def\a{$\phantom{^\ast}$}
\def\v{$\phantom{^{l}}$}
\def\pp{$\phantom{-}$}
\def\o{$\phantom{0}$}

\shorttitle{Red Supergiant Stars as Cosmic Abundance Probes}
\shortauthors{Patrick et al.}

\begin{document}


\title{Red Supergiant Stars as Cosmic Abundance Probes: \\
    KMOS Observations in NGC\,6822}


\author{L.~R.~Patrick\altaffilmark{1},
C.~J.~Evans\altaffilmark{2,1},
B.~Davies\altaffilmark{3},
R-P.~Kudritzki\altaffilmark{4,5},
J.~Z.~Gazak\altaffilmark{4},
M.~Bergemann\altaffilmark{6},
B.~Plez\altaffilmark{7},
A.~M.~N.~Ferguson\altaffilmark{1}}





\altaffiltext{1}{Institute for Astronomy, University of Edinburgh, Royal Observatory Edinburgh, Blackford Hill, Edinburgh EH9 3HJ, UK}
\altaffiltext{2}{UK Astronomy Technology Centre, Royal Observatory Edinburgh, Blackford Hill, Edinburgh EH9 3HJ, UK}
\altaffiltext{3}{Astrophysics Research Institute, Liverpool John Moores University, Liverpool Science Park ic2, 146 Brownlow Hill, Liverpool L3 5RF, UK}
\altaffiltext{4}{Institute for Astronomy, University of Hawaii, 2680 Woodlawn Drive, Honolulu, HI, 96822, USA}
\altaffiltext{5}{University Observatory Munich, Scheinerstr. 1, D-81679 Munich, Germany}
\altaffiltext{6}{Institute of Astronomy, University of Cambridge, Madingley Road, Cambridge CB3 0HA, UK}
\altaffiltext{7}{Laboratoire Univers et Particules de Montpellier, Universit\'e Montpellier 2, CNRS, F-34095 Montpellier, France}

\begin{abstract}
We present near-IR spectroscopy of red supergiant (RSG) stars in NGC\,6822, obtained with the new VLT-KMOS instrument.
From comparisons with model spectra in the $J$-band we determine the metallicity of 11 RSGs, finding a mean value of [$\bar{\rm Z}$]~=~$-$0.52\,$\pm$\,0.21 which agrees well with previous abundance studies of young stars and HII regions.
We also find an indication for a low-significance abundance gradient within the central 1\,kpc.
We compare our results with those derived from older stellar populations and investigate the difference using a simple chemical evolution model.
By comparing the physical properties determined for RSGs in NGC\,6822 with those derived using the same technique in the Galaxy and the Magellanic Clouds, we show that there appears to be no significant temperature variation of RSGs with respect to metallicity, in contrast with recent evolutionary models.
\end{abstract}



\section{Introduction}

\label{sec:introduction}
A promising new method to directly probe chemical abundances in external galaxies is with $J$-band spectroscopy of red supergiant (RSG) stars.
With their peak flux at
$\sim$1\,$\mu$m and luminosities in excess of
10$^4$\,L$_\odot$, RSGs are extremely bright in the near-IR,
making them potentially useful tracers of the chemical abundances of star-forming galaxies out to large distances.
To realise this goal,
\cite{2010MNRAS.407.1203D} outlined a technique to derive metallicities of RSGs at moderate spectral resolving power
($R\sim$3000).
This technique has recently been refined using observations of RSGs in the Magellanic Clouds
\citep{Davies-prep} and Perseus OB-1
\citep{2014ApJ...788...58G}.
Using absorption lines in the $J$-band from iron, silicon and titanium, one can estimate metallicity
([Z]~=~log\,Z/Z$_{\odot}$) as well as other stellar parameters
(effective temperature, surface gravity and microturbulence) by fitting synthetic spectra to the observations.
Owing to their intrinsic brightness,
RSGs are ideal candidates for studies of extragalactic environments in the near-IR.

To make full use of the potential of RSGs for this science, multi-object spectrographs operating in the near-IR on 8-m class telescopes are essential.
These instruments allow us to observe a large sample of RSGs in a given galaxy, at a wavelength where RSGs are brightest.
In this context, the $K$-band Multi-Object Spectrograph
\citep[KMOS;][]{2013Msngr.151...21S} at the Very Large Telescope (VLT), Chile, is a powerful facility.
KMOS will enable determination of stellar abundances for RSGs out to distances of $\sim$10\,Mpc.
Further ahead, a near-IR multi-object spectrograph on a 40-m class telescope, combined with the excellent image quality from adaptive optics,
will enable abundance estimates for individual stars in galaxies out to tens of Mpc,
a significant volume of the local universe containing entire galaxy clusters
\citep{2011A&A...527A..50E}.

Here we present KMOS observations of RSGs in the dwarf irregular galaxy NGC\,6822,
at a distance of $\sim$0.46\,Mpc
\citep[][and references therein]{2012AJ....144....4M}.
Chemical abundances have been determined for its old stellar population
\citep[e.g.][]{2001MNRAS.327..918T,2013ApJ...779..102K},
but knowledge of its recent chemical evolution and present-day abundances is
somewhat limited.
Observations of two A-type supergiants by
\cite{2001ApJ...547..765V} provided a first estimate of stellar abundances, finding
log\,(Fe/H)\,+\,12~=~7.01\,$\pm$\,0.22 and
log\,(O/H)\,+\,12~=~8.36\,$\pm$\,0.19, based on line-formation calculations for these elements
assuming local thermodynamic equilibrium (LTE).
A detailed non-LTE study for one of these objects confirmed the results finding
6.96\,$\pm$\,0.09 for iron and 8.30\,$\pm$\,0.02 for oxygen
\citep{Przybilla02}.
Compared with solar values of 7.50 and 8.69, respectively
\citep{2009ARA&A..47..481A},
this indicates abundances that are approximately one-third solar in NGC\,6822.
A study of oxygen abundances in HII regions
\citep{2006ApJ...642..813L} found a value of 8.11\,$\pm$\,0.10, confirming the low metallicity.

NGC\, 6822 is a relatively isolated Local Group galaxy, which does not seem to be associated with either M31 or the Milky Way.
It appears to have a large extended stellar halo
\citep{2002AJ....123..832L,2014ApJ...783...49H}
as well as an extended HI disk containing tidal arms and a possible HI companion
\citep{2000ApJ...537L..95D}.
The HI disk is orientated perpendicular to the distribution of old halo stars and has an associated population of blue stars
\citep{2003MNRAS.341L..39D,2003ApJ...590L..17K}.
This led \cite{2006ApJ...636L..85D} to label the system as a
\textquoteleft
polar ring galaxy\textquoteright.
A population of remote star clusters aligned with the elongated old stellar halo have been discovered
\citep{2011ApJ...738...58H,2013MNRAS.429.1039H}.
In summary, the extended structures of NGC\,6822 suggest some form of recent interaction.

In addition, there is evidence for a relatively constant star-formation history within the central 5\,kpc
\citep{2014ApJ...789..147W}
with multiple stellar populations
\citep{2006A&A...451...99B,2012A&A...540A.135S}.
This includes evidence for recent star formation in the form of a known population of massive stars, as well as a number of HII regions
\citep{2001ApJ...547..765V,2006AJ....131..343D,2009A&A...505.1027H,2012AJ....144....2L}.

In this paper we present near-IR KMOS spectroscopy of RSGs in NGC\,6822 to investigate their chemical abundances.
In Section~\ref{sec:observations} we describe the observations;
Section~\ref{sec:data_reduction} describes the data reduction and
Section~\ref{sec:results} details the derived stellar parameters and investigates the spatial distribution of the estimated metallicities in NGC\,6822.
In Section~\ref{sec:discussion} we discuss our results and
Section~\ref{sec:conclusions} concludes the paper.


\section{Observations}
\label{sec:observations}

\subsection{Target Selection} 
\label{sub:target_selection}

Our targets were selected from optical photometry
\citep{2007AJ....134.2474M}, combined with near-IR ($JHK{_s}$) photometry
\cite[for details see][]{2012A&A...540A.135S} from the Wide-Field Camera (WFCAM) on the United Kingdom Infra-Red Telescope (UKIRT).
The two catalogues were cross-matched and only sources classified as stellar in the photometry for all filters were considered.

\begin{figure}
 \includegraphics[width=9.0cm]{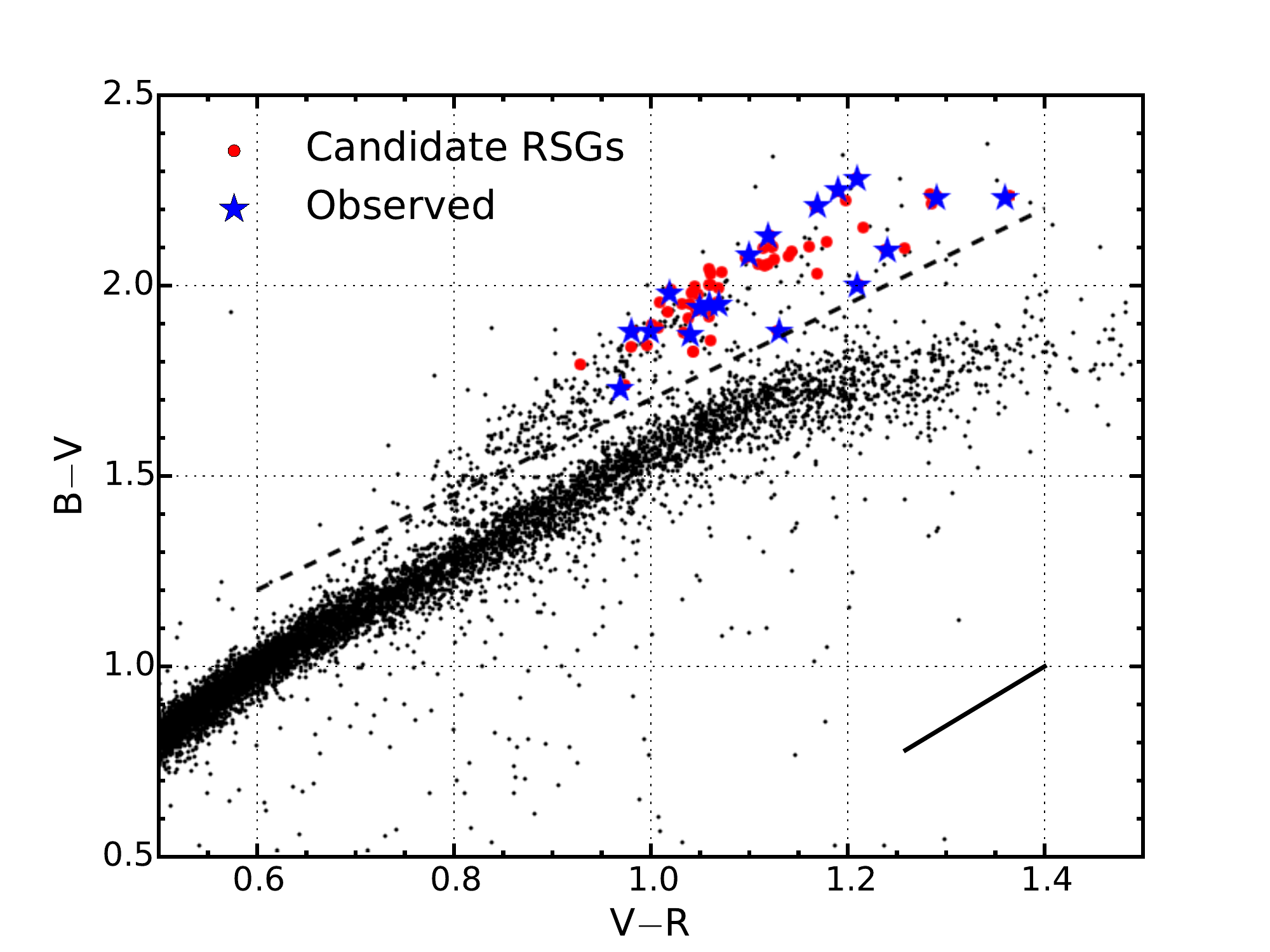}
 \caption{
          Two-colour diagram for stars with good detections in the optical and near-IR photometry in NGC\,6822.
          The black dashed line marks the selection criteria using optical colours, as defined by
          \protect\cite{2012AJ....144....2L}.
          Red circles mark all stars which satisfied our selection criteria.
          Large blue stars denote targets observed with KMOS.
          The solid black line marks the foreground reddening vector for $E(B-V)$~=~0.22
          \protect\citep{1998ApJ...500..525S}.
         }
 \label{fig:BVR}
\end{figure}

Our spectroscopic targets were selected principally based on their optical colours, as defined by
\cite{1998ApJ...501..153M} and
\cite{2012AJ....144....2L}.
Figure~\ref{fig:BVR} shows cross-matched stars, with the dividing line at
$(B$\,$-$\,$V)$~=~1.25\,$\times$\,$(V$\,$-$\,$R)$~+~0.45.
All stars redder than this line, above a given magnitude threshold and with $V$\,$-$\,$R >$~0.6 are potential RSGs.

\begin{figure}
 \includegraphics[width=9.0cm]{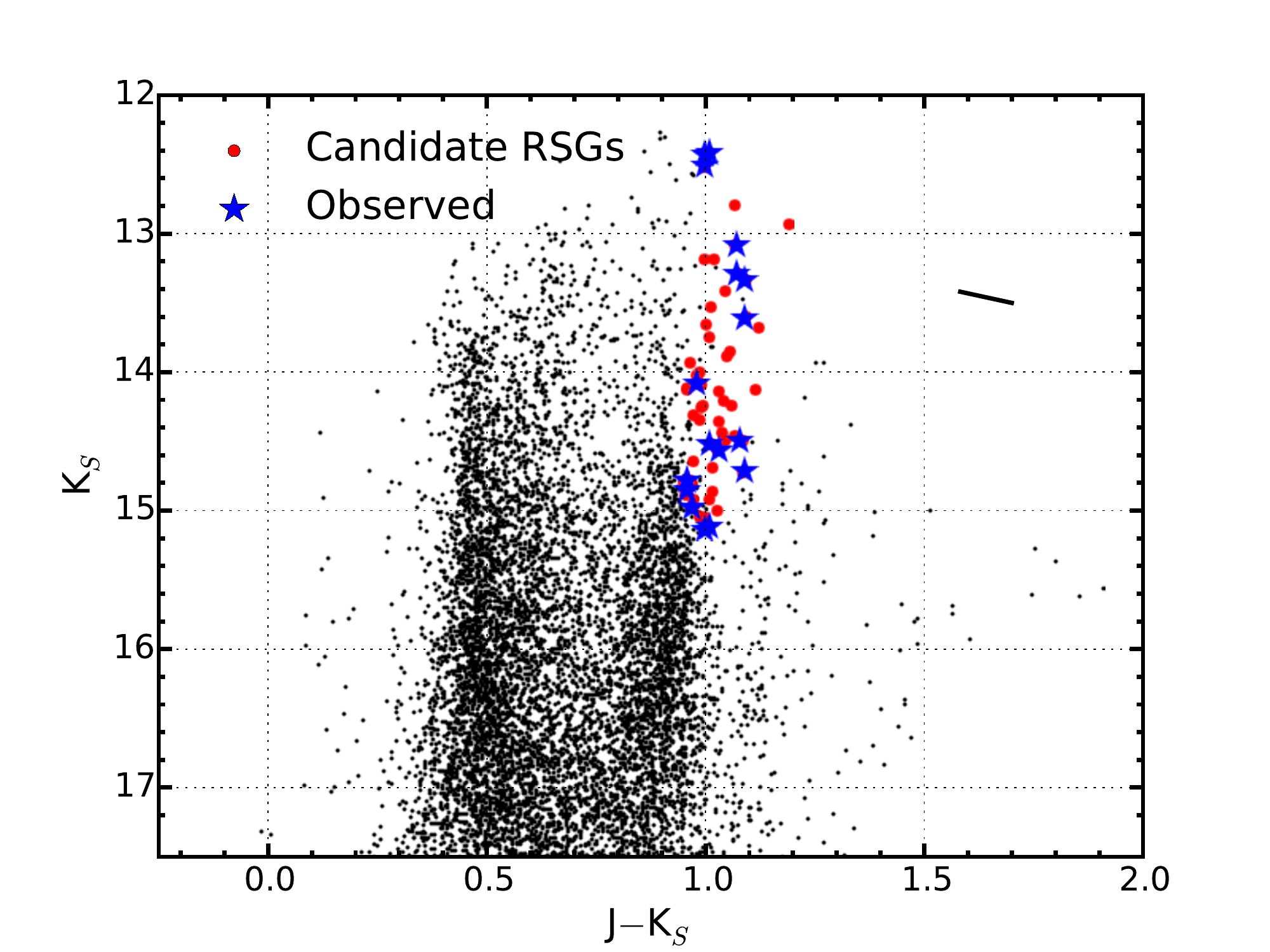}
 \caption{
          Near-IR CMD for stars classified as stellar sources in the optical and near-IR catalogues, plotted using the same symbols as Figure~\ref{fig:BVR}.
          This CMD is used to supplement the optical selection.
          The solid black line marks the foreground reddening vector for $E(B-V)$~=~0.22
          \protect\citep{1998ApJ...500..525S}.
         }
 \label{fig:JK}
\end{figure}

Distinguishing between RSGs and the most luminous stars on the asymptotic giant branch
(AGB) is difficult owing to their similar temperatures and overlapping luminosities.
Near-IR photometry can help to delineate these populations,
with RSGs located in a relatively well-defined region in the near-IR ($J-K$) colour-magnitude diagram (CMD), as discussed by
\cite{2000ApJ...542..804N}.
The near-IR CMD from the WFCAM data is shown in
Figure~\ref{fig:JK} and was used to further inform our target selection.
Employing the updated CMD criteria from
\cite{2014A&A...562A..32C}  -- modified for the distance and reddening to NGC\,6822 --
all of our potential targets from the combined optical and near-IR criteria are (notionally) RSGs.
Near-IR photometry for our targets was also given by
\cite{2013MNRAS.428.2216W} from observations with the Infrared Survey Facility;
we note that, from their analysis of the multi-epoch data,
none of our targets was identified as an AGB star.

\begin{figure}
 \includegraphics[width=9.0cm]{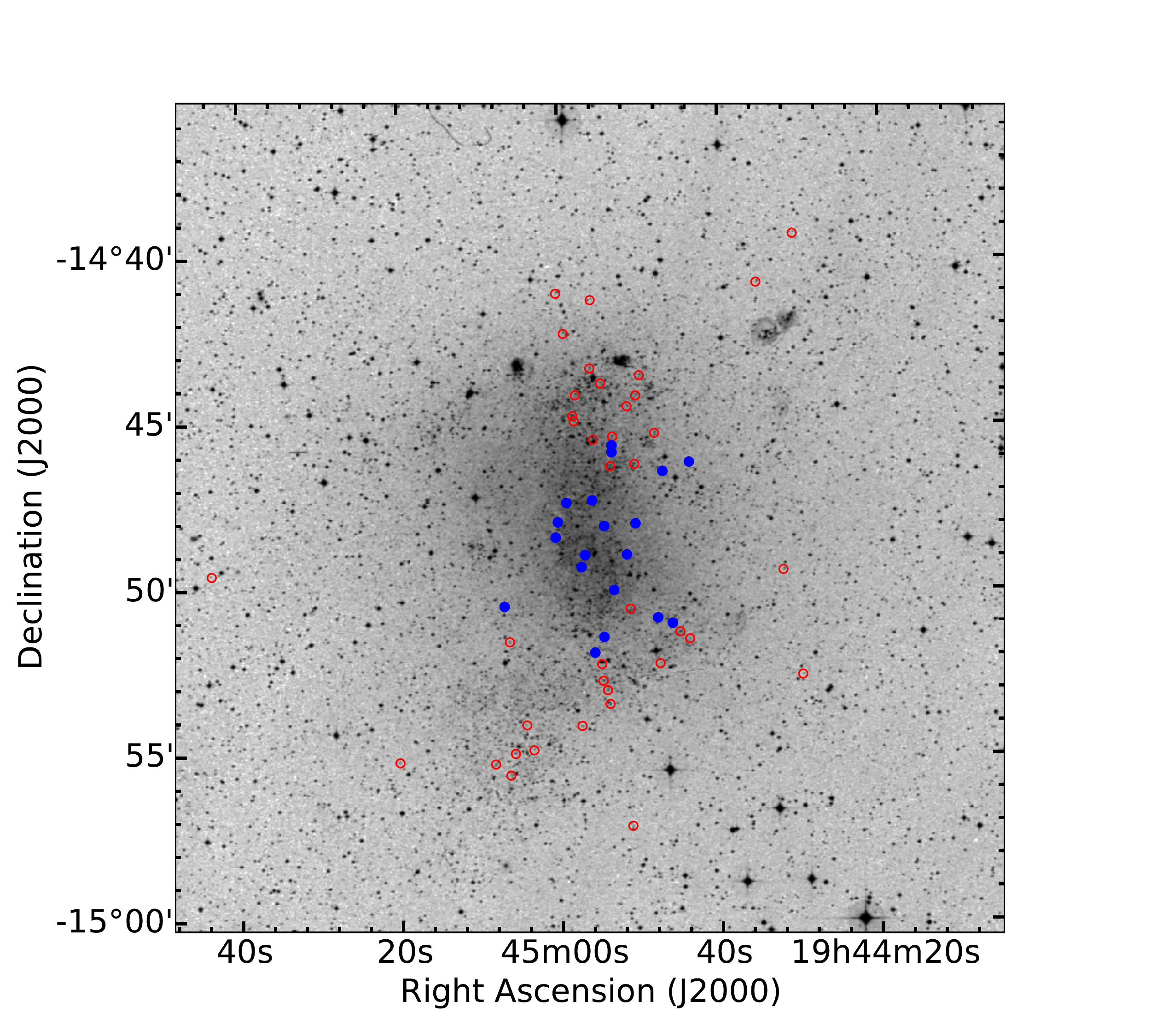}
 \caption{Spatial extent of the KMOS targets over a Digital Sky Survey (DSS) image of NGC\,6822.
          Blue filled circles indicate the locations of the observed RSGs.
          Red open circles indicate the positions of RSG candidates selected using our photometric criteria (see Section~\ref{sub:target_selection}).
          }
 \label{fig:N6822}
\end{figure}

The combined selection methods yielded 58 candidate RSGs, from which 18 stars were observed with KMOS, as shown in Figure
\ref{fig:N6822}.
The selection of the final targets was defined by the KMOS arm allocation software {\sc karma}
\citep{2008SPIE.7019E..0TW},
where the field centre was selected to maximise the number of allocated arms,
with priority given to the brightest targets.
Optical spectroscopy of eight of our observed stars, confirming them as RSGs, was presented by
\cite{2012AJ....144....2L}.

\subsection{KMOS Observations} 
\label{sub:observations}

The observations were obtained as part of the KMOS Science Verification programme on 30 June 2013 (PI: Evans, 60.A-9452(A)),
with a total exposure time of 2400\,s
(comprising 8\,$\times$\,300\,s detector integrations).
KMOS has 24 deployable integral-field units (IFUs) each of which covers an area of
2\farcs8 $\times$ 2\farcs8 within a 7\farcm2 field of view.
The 24 IFUs are split into three groups of eight, with the light from each group relayed to different spectrographs.

Offset sky frames
(0\farcm5 to the east) were interleaved between the science observations in an object (O), sky (S) sequence of:
O,\,S,\,O,\,O.
The observations were performed with the $YJ$ grating
(giving coverage from 1.02\textendash1.36\,$\mu$m);
estimates of the mean delivered resolving power for each spectrograph (obtained from the KMOS/esorex pipeline for two arc lines) are listed in Table~\ref{tb:res}.

In addition to the science observations, a standard set of KMOS calibration frames was obtained consisting of dark, flat and arc-lamp calibrations (with flats and arcs taken at six different rotator angles).
A telluric standard star was observed with the arms configured in the science positions, i.e. using the {\em KMOS\_spec\_cal\_stdstarscipatt} template in which the standard star is observed sequentially through all IFUs.
The observed standard was HIP97618, with a spectral type of B6\,III
\citep{1988mcts.book.....H}.

A summary of the observed targets is given in
Table~\ref{tb:obs-params}.
A signal-to-noise (S/N) ratio of $\gtrsim$ 100 per resolution element is required for satisfactory results from this analysis method
\citep[see][]{2014ApJ...788...58G}.
We estimated the S/N ratio of the spectra by comparing the counts in the brightest spatial pixels
(within the 1.15\textendash1.22\,$\mu$m region) of each source with the counts in equivalent spatial pixels in the corresponding sky exposures
(between the sky lines).
The S/N ratio estimated is knowingly an underestimate of the true S/N ratio achieved.


\begin{table*}
\caption{Measured velocity resolution and resolving power ($R$) across each detector\label{tb:res}}
\scriptsize
\begin{center}
\begin{tabular}{crcccc}
\hline
\hline
Detector & IFUs & \multicolumn{2}{c}{Ne\,\lam1.17700\,$\mu$m}
            & \multicolumn{2}{c}{Ar\,\lam1.21430\,$\mu$m} \\
 & & FWHM (\kms) & $R$ & FWHM (\kms) & $R$ \\
  \hline
1 & 1\textendash8 &  \a88.04\,$\pm$\,2.67 & 3408\,$\pm$\,103 &
                     \a85.45\,$\pm$\,2.67 & 3511\,$\pm$\,110 \\
2 & 9\textendash16 & \a82.83\,$\pm$\,2.48 & 3622\,$\pm$\,108 &
                     \a80.30\,$\pm$\,3.05 & 3736\,$\pm$\,142 \\
3 & 17\textendash24 & 103.23\,$\pm$\,2.73 & 2906\,$\pm$\,77\a &
                      101.25\,$\pm$\,2.99 & 2963\,$\pm$\,87\a \\
\hline
\end{tabular}
\end{center}
\end{table*}

\begin{table*}
\caption{
        Summary of VLT-KMOS targets in NGC\,6822\label{tb:obs-params}
        }
\scriptsize
\begin{center}
\begin{tabular}{lrcccccccccl}
 \hline
 \hline
ID & S/N & $\alpha$ (J2000) & $\delta$ (J2000) & $B$ & $V$ & $R$ & $J$ & $H$ & $K_{\rm s}$ & RV (\kms) & Notes \\
 \hline
NGC6822-RSG01 & 223 &   19:44:43.81  &  $-$14:46:10.7  &  20.83  &  18.59  &  17.23  &  14.16  &  13.37  &  13.09  &   $-$69.9\,$\pm$\,3.7 & Sample\\
NGC6822-RSG02 & 120 &   19:44:45.98  &  $-$14:51:02.4  &  20.91  &  18.96  &  17.89  &  15.53  &  14.72  &  14.52  &   $-$66.4\,$\pm$\,6.6 & Sample\\
NGC6822-RSG03 &  94 &   19:44:47.13  &  $-$14:46:27.1  &  21.30  &  19.41  &  18.41  &  16.13  &  15.35  &  15.12  &   $-$43.0\,$\pm$\,6.0  \\
NGC6822-RSG04 & 211 &   19:44:47.81  &  $-$14:50:52.5  &  20.74  &  18.51  &  17.22  &  14.37  &  13.58  &  13.30  &   $-$64.5\,$\pm$\,3.1 & LM12 (M1), Sample \\
NGC6822-RSG05 & 104 &   19:44:50.54  &  $-$14:48:01.6  &  20.83  &  18.95  &  17.97  &  15.75  &  14.98  &  14.79  & \a$-$93.4\,$\pm$\,17.1 \\
NGC6822-RSG06 & 105 &   19:44:51.64  &  $-$14:48:58.0  &  21.33  &  19.45  &  18.32  &  15.81  &  14.95  &  14.72  & \a\a\a\ldots \\
NGC6822-RSG07 & 145 &   19:44:53.46  &  $-$14:45:52.6  &  20.36  &  18.43  &  17.38  &  15.06  &  14.30  &  14.08  &   $-$72.6\,$\pm$\,3.5 & LM12 (M4.5), Sample \\
NGC6822-RSG08 & 103 &   19:44:53.46  &  $-$14:45:40.1  &  20.88  &  19.14  &  18.17  &  15.95  &  15.16  &  14.98  &   $-$73.0\,$\pm$\,5.8 & LM12 (K5), Sample \\
NGC6822-RSG09 & 201 &   19:44:54.46  &  $-$14:48:06.2  &  20.56  &  18.56  &  17.35  &  14.43  &  13.67  &  13.34  &   $-$79.9\,$\pm$\,3.7 & LM12 (M1), Sample\\
NGC6822-RSG10 & 302 &   19:44:54.54  &  $-$14:51:27.1  &  19.29  &  17.05  &  15.86  &  13.43  &  12.66  &  12.42  &   $-$57.2\,$\pm$\,4.7 & LM12 (M0), Sample \\
NGC6822-RSG11 & 327 &   19:44:55.70  &  $-$14:51:55.4  &  19.11  &  16.91  &  15.74  &  13.43  &  12.70  &  12.43  &   $-$67.9\,$\pm$\,3.1 & LM12 (M0), Sample \\
NGC6822-RSG12 & 100 &   19:44:55.93  &  $-$14:47:19.6  &  21.43  &  19.56  &  18.52  &  16.14  &  15.33  &  15.14  &   $-$61.2\,$\pm$\,6.4 & LM12 (K5) \\
NGC6822-RSG13 & 106 &   19:44:56.86  &  $-$14:48:58.5  &  21.05  &  19.06  &  18.04  &  15.81  &  15.05  &  14.85  & \a$-$55.0\,$\pm$\,10.0 \\
NGC6822-RSG14 & 284 &   19:44:57.31  &  $-$14:49:20.2  &  19.69  &  17.41  &  16.20  &  13.52  &  12.76  &  12.52  &   $-$52.7\,$\pm$\,4.4  & LM12 (M1), Sample \\
NGC6822-RSG15 & 124 &   19:44:59.14  &  $-$14:47:23.9  &  21.30  &  19.17  &  18.05  &  15.58  &  14.74  &  14.50  & \a$-$74.2\,$\pm$\,13.4  \\
NGC6822-RSG16 & 107 &   19:45:00.24  &  $-$14:47:58.9  &  21.27  &  19.20  &  18.10  &  15.60  &  14.80  &  14.57  &   $-$68.8\,$\pm$\,4.5 \\
NGC6822-RSG17 & 167 &   19:45:00.53  &  $-$14:48:26.5  &  20.84  &  18.75  &  17.51  &  14.70  &  13.86  &  13.61  &   $-$62.1\,$\pm$\,4.0 & Sample\\
NGC6822-RSG18 & 104 &   19:45:06.98  &  $-$14:50:31.1  &  21.06  &  19.12  &  18.06  &  15.74  &  14.94  &  14.78  &   $-$87.8\,$\pm$\,9.4 & Sample\\
\hline
\end{tabular}
\end{center}
\tablecomments{Optical data from
\protect\cite{2007AJ....134.2474M}, with typical photometric uncertainty 0.016, 0.006, 0.010 in $B, V$ and $R$ bands, respectively.
Near-IR data from the UKIRT survey
\protect\cite[see][for details]{2012A&A...540A.135S}, with typical errors 0.015, 0.010, 0.012, in $J, H$ and $K$ bands, respectively.
Targets observed by
\protect\cite{2012AJ....144....2L}
are indicated by \textquoteleft
LM12\textquoteright~ in the final column (with their spectral classifications in parentheses).
Targets used for abundance analysis are indicated by the comment
\textquoteleft Sample\textquoteright
.}
\end{table*}


\section{Data Reduction} 
\label{sec:data_reduction}

The observations were reduced using the recipes provided by the Software Package for Astronomical Reduction with KMOS
\citep[SPARK;][]{2013A&A...558A..56D}.
The standard KMOS/esorex routines were used to calibrate and reconstruct the science and standard-star data cubes as outlined by
\cite{2013A&A...558A..56D}.
Sky subtraction was performed using the standard KMOS recipes and telluric correction was performed using two different strategies.
Throughout the following analysis all spectra have been extracted from their respective data cubes using a consistent method (i.e. the optimal extractions within the pipeline).

\subsection{Three-arm vs. 24-arm Telluric Correction} 
\label{sub:three_arm_vs_24_arm_telluric_correction}

The default template for telluric observations with KMOS is to observe a standard star in one IFU in each of the three spectrographs.
However, there is an alternative template which allows users to observe a standard star in each of the 24 IFUs.
This strategy should provide an optimum telluric correction for the KMOS IFUs but reduces observing efficiency.

A comparison between the two methods in the $H$-band was given by
\cite{2013A&A...558A..56D},
who concluded that using the more efficient three-arm method was suitable for most science purposes.
However, an equivalent analysis in the $YJ$-band was not available.
To determine if the more rigorous telluric approach is required for our analysis,
we observed a telluric standard star (HIP97618) in each of the 24 IFUs.
This gave us the data to investigate both of the telluric correction methods and to directly compare the results.

\begin{figure*}
 \begin{center}
 \includegraphics[width=12.0cm]{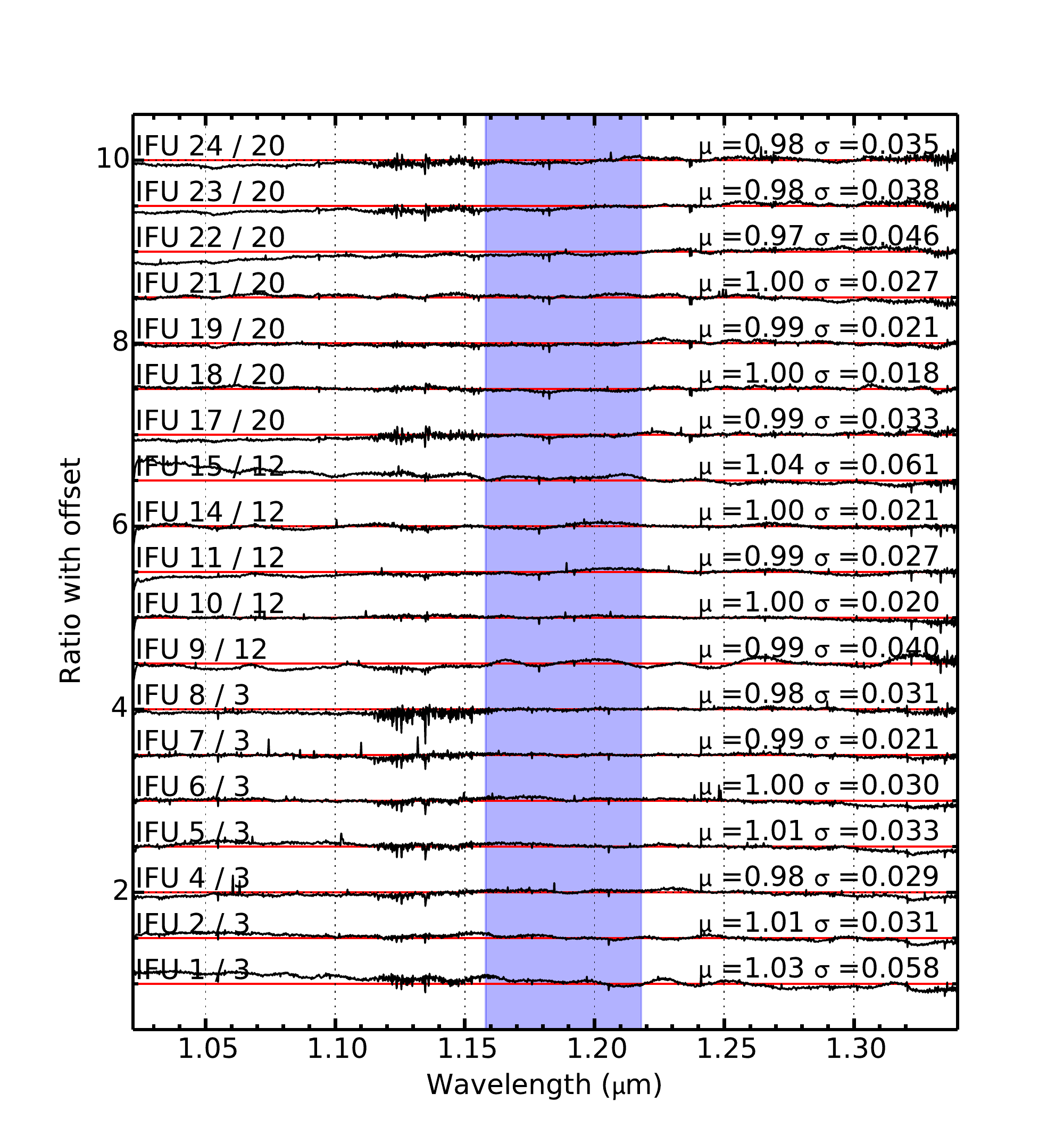}
 \caption{
    Comparison of $J$-band spectra of the same standard star in each IFU.
    The ratio of each spectrum compared with that from the IFU used in the three-arm telluric method is shown,
    with their mean and standard deviation ($\mu$ and $\sigma$ respectively).
    Red lines indicate $\mu$~=~1.0, $\sigma$~=~0.0 for each ratio.
    The blue shaded area signifies the region used in our analysis,
    within which the discrepancies between the IFUs are generally small.
    This is reflected in the standard deviation values when only considering this region.
    (IFUs 13 and 16 are omitted as no data were taken with these IFUs.) \label{fig:IFU_compare}
          }
 \end{center}
\end{figure*}

We first compared the standard-star spectrum in each IFU with that used by the pipeline routines for the three-arm template in each of the spectrographs.
Figure~\ref{fig:IFU_compare} shows the differences between the standard-star spectra across the IFUs,
where the differences in the $YJ$-band are comparable to those in the $H$-band
\cite[cf. Fig.~7 from][]{2013A&A...558A..56D}.
The qualitative agreement between the IFUs in our region of interest (1.15\textendash1.22\,$\mu$m) is generally very good.

To quantify the difference the two telluric methods would make to our analysis,
we performed the steps described in
Section~\ref{sub:ngc6822_telluric_correction} for both templates.
We then used the two sets of reduced science data
(reduced with both methods of the telluric correction) to compute stellar parameters for our targets.
The results of this comparison are detailed in Section
\ref{sub:telluric_comparison}.


\subsection{Telluric Correction Implementation} 
\label{sub:ngc6822_telluric_correction}

To improve the accuracy of the telluric correction,
for both methods mentioned above,
we implemented additional recipes beyond those of the KMOS/esorex pipeline.
These recipes were employed to account for two effects which could potentially degrade the quality of the telluric correction.
The first corrects for any potential shift in wavelength between each science spectrum and its associated telluric standard.
The most effective way to implement this is to cross-correlate each pair of science and telluric-standard spectra.
Any shift between the two is then applied to the telluric standard using a cubic-spline interpolation routine.

The second correction applied is a simple spectral scaling algorithm.
This routine corrects for differences in line intensity of the most prominent features common to both the telluric and the science spectra.
To find the optimal scaling parameter the following formula is used:

\begin{equation} \label{eq:shiftandres}
T_{2} = (T_{1} + c) / (T_{1} - c),
\end{equation}

\noindent where $T_{2}$ is the corrected telluric-standard spectrum,
$T_{1}$ is the initial telluric standard spectrum and $c$ is the scaling parameter.

To determine the required scaling,
telluric spectra are computed for $-$0.5~$<c<$~0.5, in increments of 0.02
(where a perfect value, i.e. no difference in line strength, would be $c$~=~0).
Each telluric spectrum is used to correct the science data and the standard deviation of the counts across the spectral region is computed for each corrected spectrum.
The minimum value of the standard-deviation matrix defines the optimum scaling.
For this algorithm, only the region of interest for our analysis is considered
(i.e. 1.15\textendash1.22\,$\mu$m).

The final set of telluric-standard spectra from the KMOS/esorex reductions were modified using these additional routines and were then used to correct the science observations for the effects of the Earth's atmosphere.

As an alternative to observing telluric standard stars, a new telluric correction package, {\sc molecfit}, allows one to calculate a telluric spectrum based on atmospheric modelling.
Briefly, the software uses a reference atmospheric profile to estimate the true profile for the time and location of the science observation.
This model is then used to create a telluric spectrum which can be used to correct the observations.

This software has been shown to work well on a variety of VLT instruments
(Smette et al. submitted) and has been rigorously tested using X-shooter spectra
(Kausch et al. submitted).
However, the package has yet to be tested thoroughly on lower resolution observations such as those from KMOS.
Our first tests appear encouraging; however, pending further characterisation of the KMOS data cubes
(e.g. small variations in spectral resolving power leading to sky residuals,
see Section~\ref{sub:sky_subtraction}),
we will investigate the potential of the {\sc molecfit} package in future papers of this series.


\subsection{Sky Subtraction} 
\label{sub:sky_subtraction}

Initial inspection of the extracted stellar spectra revealed minor residuals from the sky subtraction process.
Reducing these cases with the \textquoteleft sky\_tweak\textquoteright
~option within the KMOS/esorex reduction pipeline was ineffective to improve the subtraction of these features.
Any residual sky features could potentially influence our results by perturbing the continuum placement within the model fits, which is an important aspect of the fitting process
\citep[see][for more discussion]{2014ApJ...788...58G,Davies-prep}.
Thus, pending a more rigorous treatment of the data
(e.g. to take into account the changing spectral resolution across the array),
we exclude objects showing sky residuals from our analysis.
Of the 18 observed targets, 11 were used to derive stellar parameters
(as indicated in Table~\ref{tb:obs-params}).

\subsection{Stellar Radial Velocities} 
\label{sub:RVs}

Radial velocities for each target are listed in Table~\ref{tb:obs-params}.
To check the accuracy of the wavelength solution provided by the data reduction pipeline,
this solution is cross-correlated with the wavelength solution of a spectrum of the Earth's telluric features\footnotemark[1].
In general, this is a small correction.
The science spectra are then telluric corrected in the manner described above.

\footnotetext[1]{Retrieved from http://www.eso.org/sci/facilities/paranal/
decommissioned/isaac/tools/spectroscopic\_standards.html}

Radial velocities are estimated by measuring the peak of the cross-correlation function between the telluric-corrected science spectra
(taking the spectrum from the brightest spatial pixel)
and an appropriate synthetic RSG spectrum.
The errors are calculated by taking the dispersion of each line used to derive the radial velocities scaled by the number of lines used, following
\cite{2015ApJ...798...23L}.
A robust radial velocity could not be derived for one of our candidates
(NGC6822-RSG06) owing to strong residual sky features; therefore,
it is not considered further.

\begin{figure}
\includegraphics[width=9.0cm]{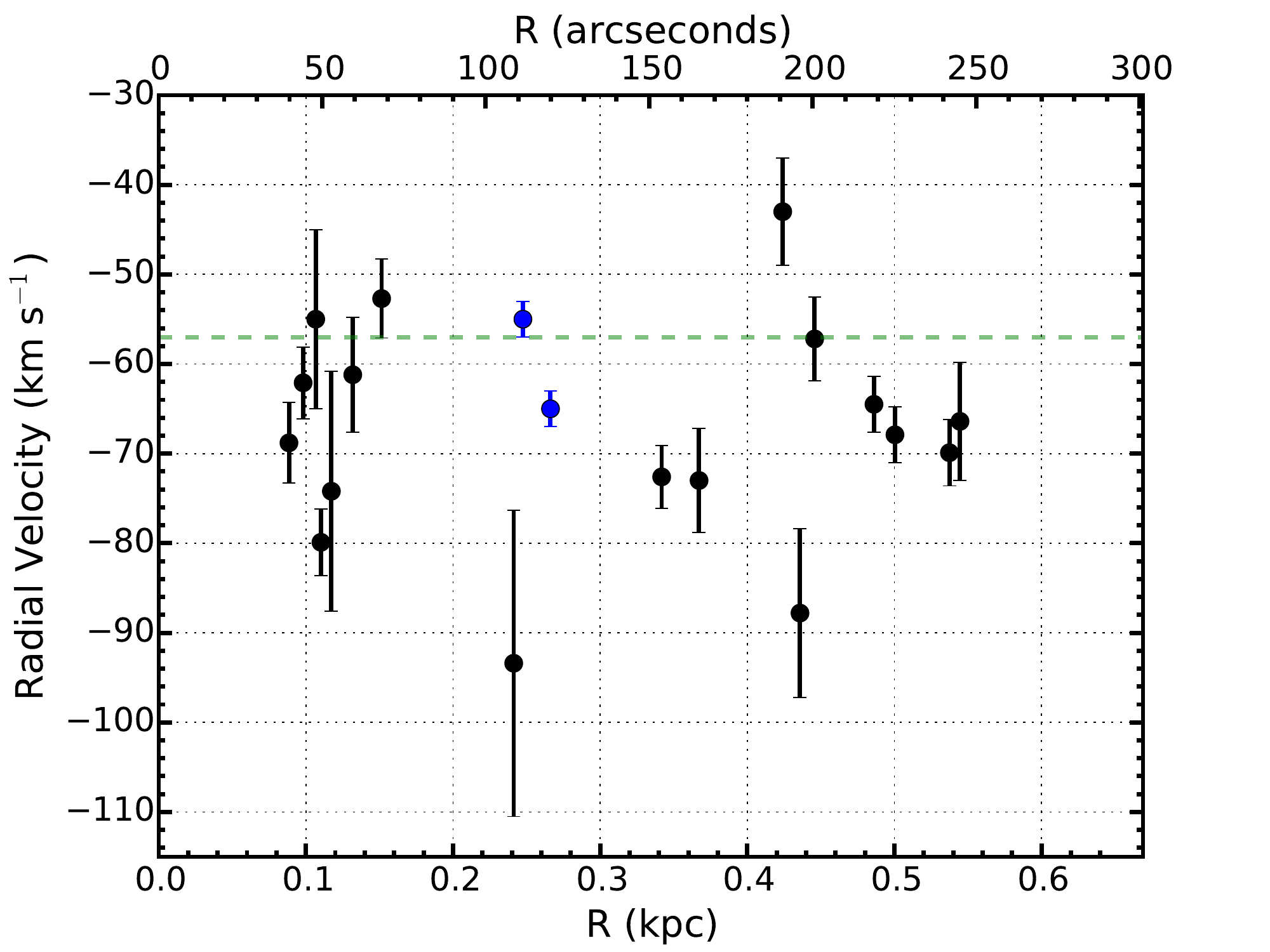}
\caption{
Radial velocities of targets shown against their distance from the galaxy centre.
The average radial velocity for the sample is $-68\pm12$\,\kms.
The green dashed line indicates the H\,I systemic velocity
\protect\citep[$-57\pm2$\,\kms;][]{2004AJ....128...16K}.
The radial velocities of two A-type supergiants from
\protect\cite{2001ApJ...547..765V} are shown in blue.
        }
\label{fig:RvsRV}
\end{figure}

Radial velocity estimates are shown as a function of distance to the centre of NGC\,6822 in Figure~\ref{fig:RvsRV}.
The average radial velocity for our targets is $-$68\,$\pm$\,12\,\kms,
in good agreement with the systemic radial velocity of the H\,I disk
\citep[$-$57\,$\pm$\,2\,\kms;][]{2004AJ....128...16K}.
Our radial velocities also agree with estimates for the two A-type supergiants from
\cite{2001ApJ...547..765V}.
This result confirms that our candidates are NGC\,6822 members.

\section{Results} 
\label{sec:results}

\begin{figure*}
 \begin{center}
\includegraphics[width=16cm]{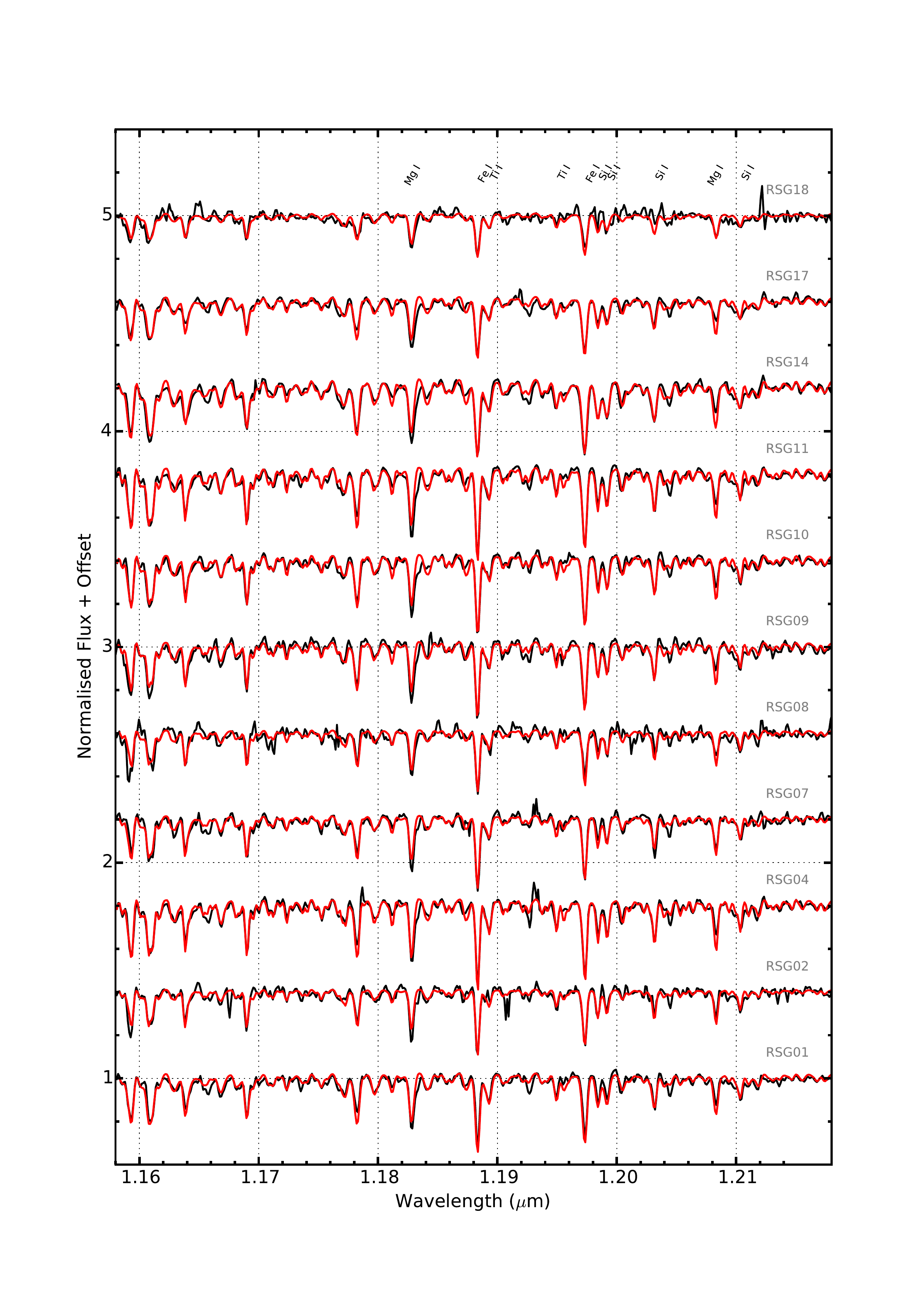}
\caption{KMOS spectra of the NGC\,6822 RSGs and their associated best-fit model spectra
(black and red lines, respectively).
The lines used for the analysis from left to right by species are:
Fe\,I$\lambda\lambda$1.188285,
1.197305,
Si\,I$\lambda\lambda$1.198419,
1.199157,
1.203151,
1.210353,
Ti\,I$\lambda\lambda$1.189289,
1.194954.
The two strong Mg\,I lines are also labelled, but are not used in the fits
(see Section~\ref{sec:results}).
         }
\label{fig:model_fits}
\end{center}
\end{figure*}

Stellar parameters
(metallicity, effective temperature, surface gravity and microturbulence)
have been derived using the $J$-band analysis technique described by
\cite{2010MNRAS.407.1203D} and demonstrated by
\cite{2014ApJ...788...58G} and
\cite{Davies-prep}.
To estimate physical parameters this technique uses a grid of synthetic spectra to fit observational data,
in which the models are degraded to the resolution of the observed spectra
(Table~\ref{tb:res}).
Model atmospheres were generated using the {\sc marcs} code
\citep{2008A&A...486..951G} where the range of parameters is defined in
Table~\ref{tb:mod_range}.
The precision of the models is increased by including departures from LTE in some of the strongest Fe, Ti and Si atomic lines
\citep{2012ApJ...751..156B,2013ApJ...764..115B}.
The two strong magnesium lines in our diagnostic spectral region are excluded from the analysis at present as these lines are known to be affected strongly by non-LTE effects
(see Figure~\ref{fig:model_fits}, where the two Mg\,I lines are systematically under- and over-estimated).
The non-LTE effects on the formation of the Mg I lines will be explored by
Bergemann et al. (submitted).

\begin{table}
\caption{
Model grid used for analysis\label{tb:mod_range}
         }
\scriptsize
\begin{center}
\begin{tabular}{lccc}
 \hline
 \hline
  Model Parameter & Min. & Max. & Step size \\
 \hline
T$_{\rm eff}$ (K)    & 3400 & 4000  & 100 \\
                     & 4000 & 4400  & 200 \\
$[$Z$]$ (dex)   & $-$1.50   & 1.00  & 0.25\\
log\,$g$ (cgs)  & $-$1.0\a  & 1.0\a & 0.5\v \\
 $\xi$ (\kms)   & \pp1.0\a  & 6.0\a & 1.0\v\\
 \hline
\end{tabular}
\end{center}
\end{table}

\subsection{Telluric Comparison} 
\label{sub:telluric_comparison}

\begin{figure*}
 \begin{center}$
  \centering
  \begin{array}{cc}
  \includegraphics[width=9.0cm]{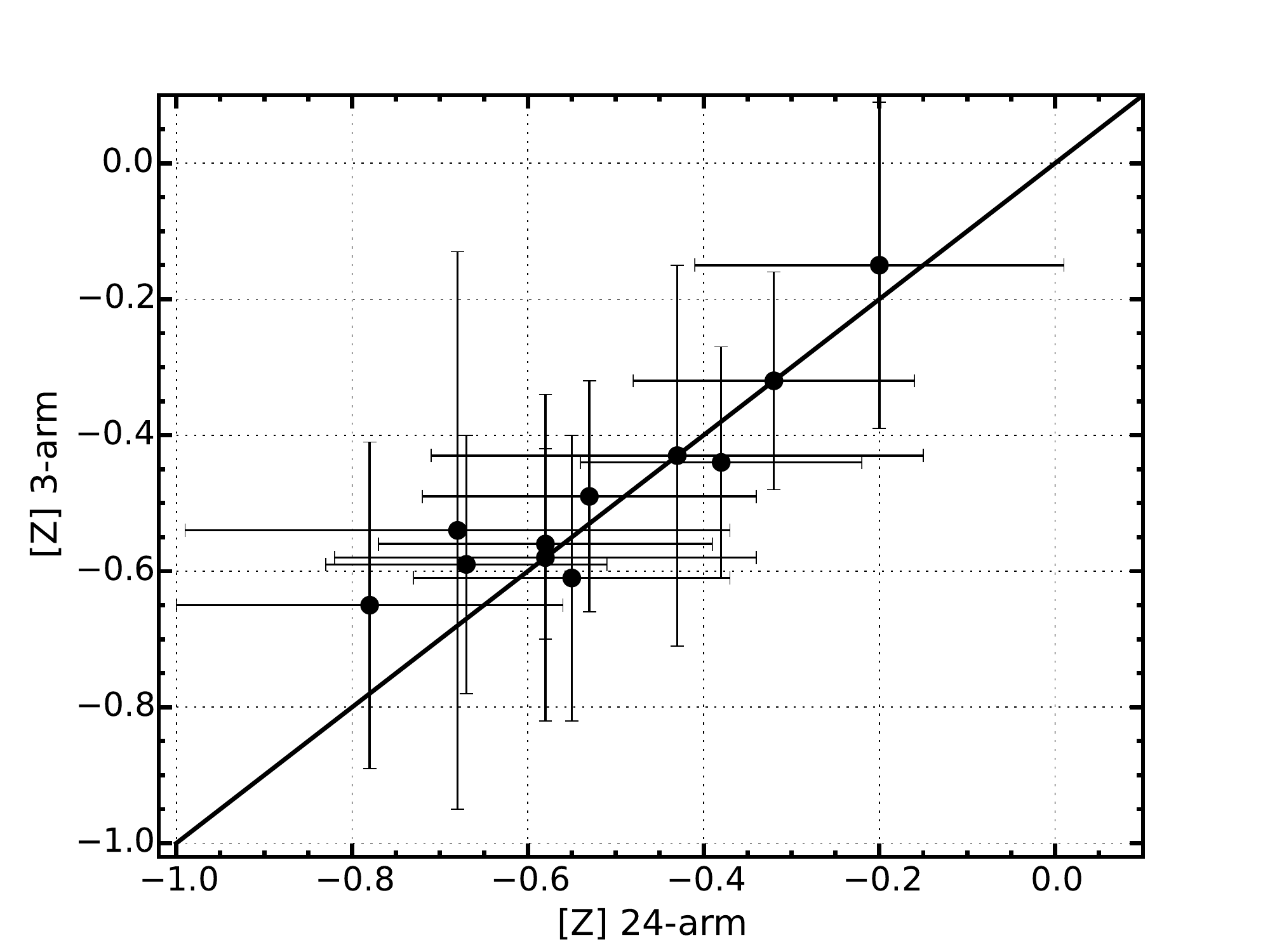} &
  \includegraphics[width=9.0cm]{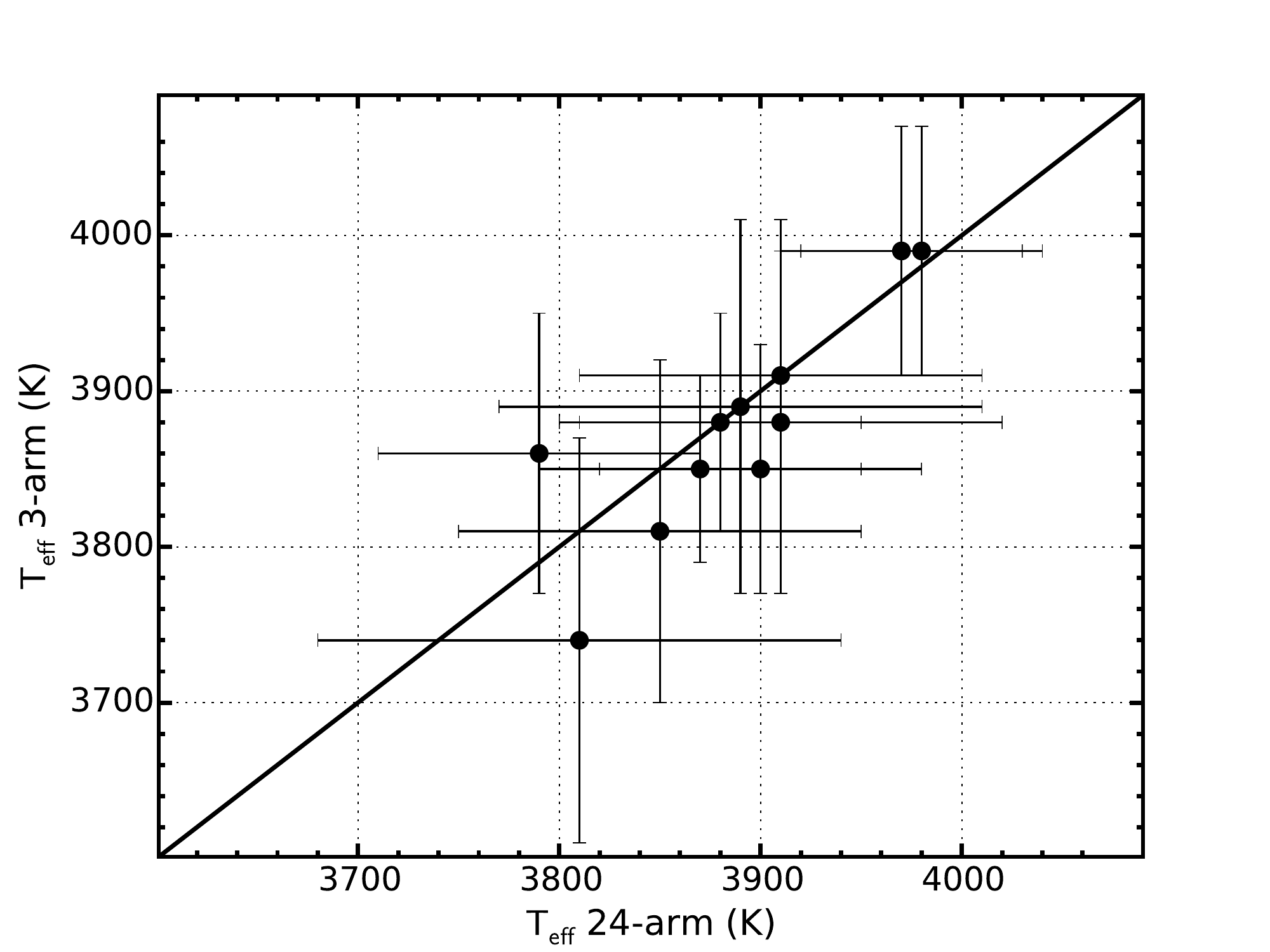} \\
  \includegraphics[width=9.0cm]{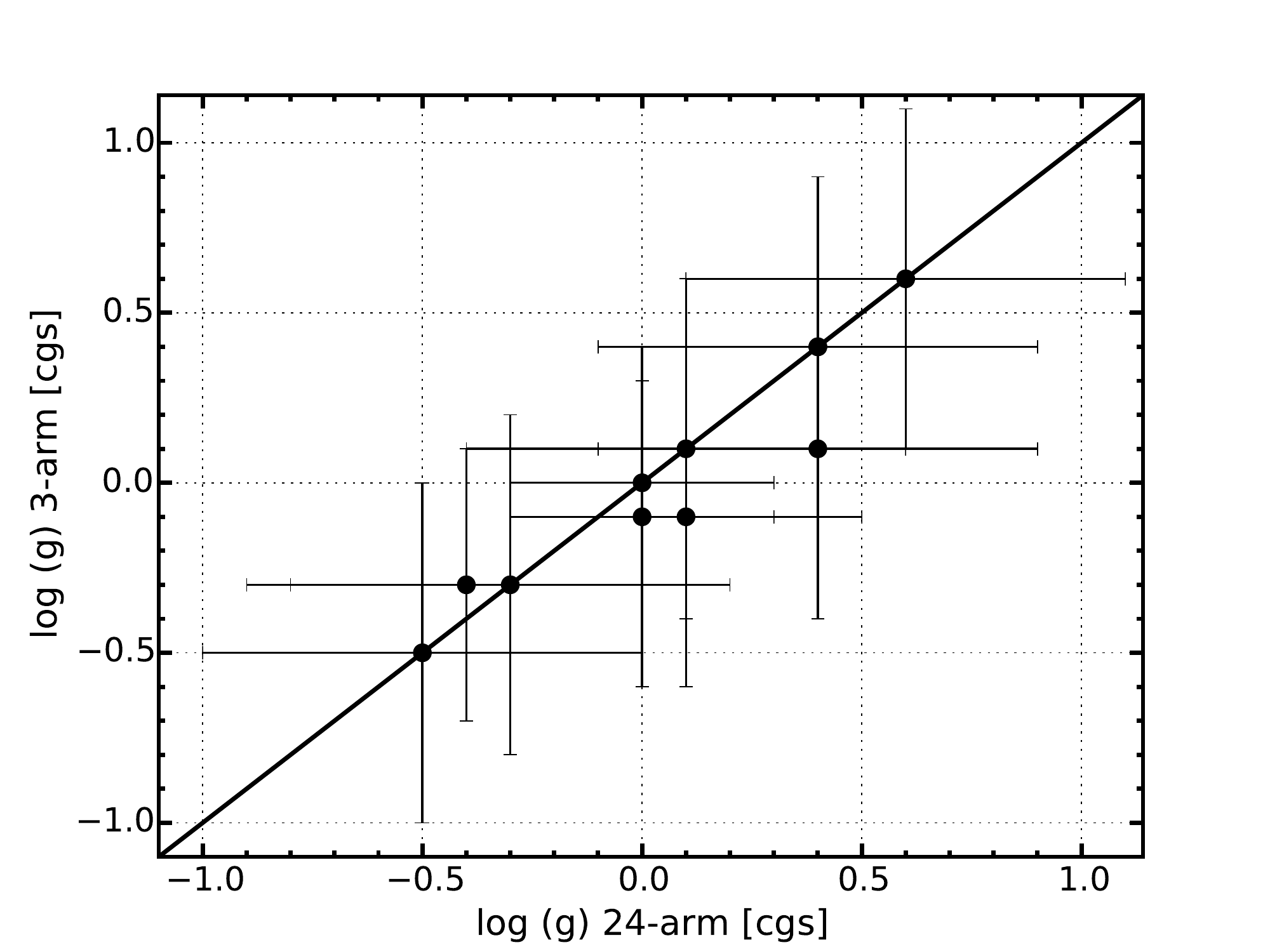} &
  \includegraphics[width=9.0cm]{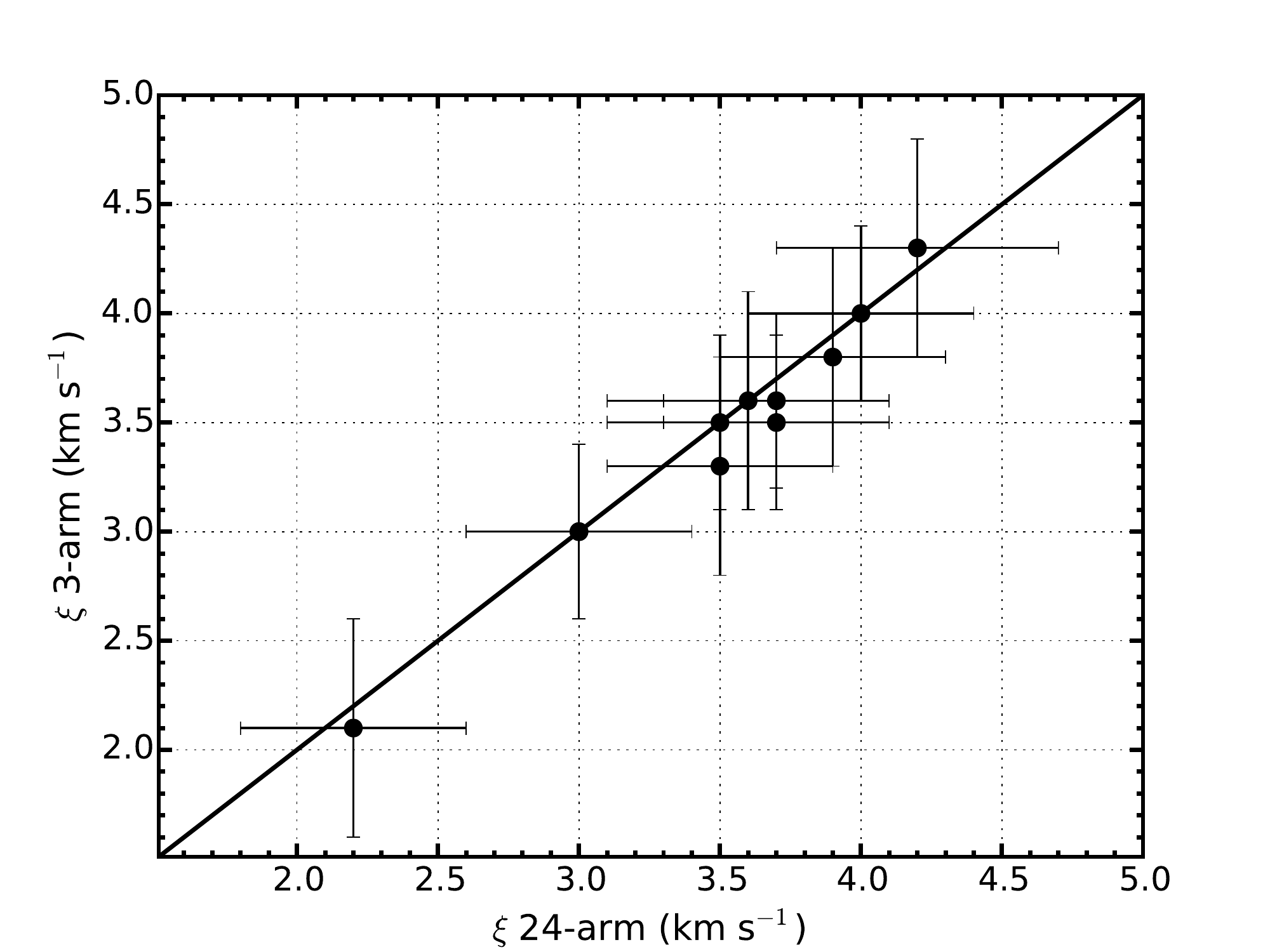} \\
  \end{array}$
 \end{center}
 \caption{
            Comparison of the model parameters using the two different telluric methods.
            In each panel, the x-axis represents stellar parameters derived using the three-arm telluric
            method and the y-axis represents those derived using the 24-arm telluric method.
            Top left: metallicity ([Z]), mean difference
            $<\Delta[$Z$]>$~=~0.04\,$\pm$\,0.07.
            Top right: effective temperature (T$_{\rm eff}$), mean difference
            $<\Delta $T$_{\rm eff}>$~=~$-$14\,$\pm$\,42.
            Bottom left: surface gravity (log\,$g$), mean difference
            $<\Delta$ log\,$g>$~=~$-$0.06\,$\pm$\,0.12.
            Bottom right: microturbulence ($\xi$), mean difference
            $<\Delta \xi>$~=~$-$0.1\,$\pm$\,0.1.
            In all cases, the distributions are statistically consistent with a one-to-one ratio (black lines).
          }
 \label{fig:3vs24AT}
\end{figure*}

We used these Science Verification data to determine which of the two telluric standard methods is most appropriate for our analysis.
Table~\ref{tb:stellar-params} details the stellar parameters derived for each target using both telluric methods, and these parameters are compared in
Figure~\ref{fig:3vs24AT}.
The mean difference in metallicity from the two methods is
$\Delta [$Z$]$~=~0.04\,$\pm$\,0.07.
Therefore, for our analysis, there is no significant difference between the two telluric approaches.

\begin{table*}
\begin{center}
\caption{
Fit parameters for reductions using the two different telluric methods.
\label{tb:stellar-params}
         }
\scriptsize
\begin{tabular}{lc cccc c cccc}
 \hline
 \hline
  Target  & IFU &  \multicolumn{4}{c}{24-arm Telluric} & \multicolumn{4}{c}{three-arm Telluric}\\
  \cline{3-6}  \cline{8-11}
 &  & T$_{\rm eff}$ (K) & log\,$g$ & $\xi$ (\kms) & [Z] & & T$_{\rm eff}$ (K) & log\,$g$ & $\xi$ (\kms) & [Z]\\
  \hline
NGC6822-RSG01 & 6 & 3790\,$\pm$\,80\o & $-$0.0\,$\pm$\,0.3 & 3.5\,$\pm$\,0.4 & $-$0.55\,$\pm$\,0.18 & & 3860\,$\pm$\,90\o & $-$0.1\,$\pm$\,0.5 &  3.5\,$\pm$\,0.4 & $-$0.61\,$\pm$\,0.21 \\
NGC6822-RSG02 & 11& 3850\,$\pm$\,100  & \pp0.4\,$\pm$\,0.5 & 3.5\,$\pm$\,0.4 & $-$0.78\,$\pm$\,0.22 & & 3810\,$\pm$\,110  & \pp0.4\,$\pm$\,0.5 &  3.3\,$\pm$\,0.5 & $-$0.65\,$\pm$\,0.24 \\
NGC6822-RSG04 & 12& 3880\,$\pm$\,70\o & \pp0.0\,$\pm$\,0.3 & 4.0\,$\pm$\,0.4 & $-$0.32\,$\pm$\,0.16 & & 3880\,$\pm$\,70\o & \pp0.0\,$\pm$\,0.3 &  4.0\,$\pm$\,0.4 & $-$0.32\,$\pm$\,0.16 \\
NGC6822-RSG07 & 2 & 3970\,$\pm$\,60\o & \pp0.4\,$\pm$\,0.5 & 3.9\,$\pm$\,0.4 & $-$0.58\,$\pm$\,0.19 & & 3990\,$\pm$\,80\o & \pp0.1\,$\pm$\,0.5 &  3.8\,$\pm$\,0.5 & $-$0.56\,$\pm$\,0.14 \\
NGC6822-RSG08 & 3 & 3910\,$\pm$\,100  & \pp0.6\,$\pm$\,0.5 & 3.0\,$\pm$\,0.4 & $-$0.58\,$\pm$\,0.24 & & 3910\,$\pm$\,100  & \pp0.6\,$\pm$\,0.5 &  3.0\,$\pm$\,0.4 & $-$0.58\,$\pm$\,0.24 \\
NGC6822-RSG09 & 4 & 3980\,$\pm$\,60\o & \pp0.1\,$\pm$\,0.4 & 3.7\,$\pm$\,0.4 & $-$0.38\,$\pm$\,0.16 & & 3990\,$\pm$\,80\o & $-$0.1\,$\pm$\,0.5 &  3.6\,$\pm$\,0.4 & $-$0.44\,$\pm$\,0.17 \\
NGC6822-RSG10 & 14& 3900\,$\pm$\,80\o & $-$0.3\,$\pm$\,0.5 & 3.7\,$\pm$\,0.4 & $-$0.67\,$\pm$\,0.16 & & 3850\,$\pm$\,80\o & $-$0.3\,$\pm$\,0.5 &  3.5\,$\pm$\,0.4 & $-$0.59\,$\pm$\,0.19 \\
NGC6822-RSG11 & 15& 3870\,$\pm$\,80\o & $-$0.4\,$\pm$\,0.5 & 4.2\,$\pm$\,0.5 & $-$0.53\,$\pm$\,0.19 & & 3850\,$\pm$\,60\o & $-$0.3\,$\pm$\,0.4 &  4.3\,$\pm$\,0.5 & $-$0.49\,$\pm$\,0.17 \\
NGC6822-RSG14 & 17& 3910\,$\pm$\,110  & $-$0.5\,$\pm$\,0.5 & 3.6\,$\pm$\,0.5 & $-$0.20\,$\pm$\,0.21 & & 3880\,$\pm$\,110  & $-$0.5\,$\pm$\,0.5 &  3.6\,$\pm$\,0.5 & $-$0.15\,$\pm$\,0.24 \\
NGC6822-RSG17 & 21& 3890\,$\pm$\,120  & \pp0.1\,$\pm$\,0.5 & 3.0\,$\pm$\,0.4 & $-$0.43\,$\pm$\,0.28 & & 3890\,$\pm$\,120  & \pp0.1\,$\pm$\,0.5 &  3.0\,$\pm$\,0.4 & $-$0.43\,$\pm$\,0.28 \\
NGC6822-RSG18 & 18& 3810\,$\pm$\,130  & \pp0.4\,$\pm$\,0.5 & 2.2\,$\pm$\,0.4 & $-$0.68\,$\pm$\,0.31 & & 3740\,$\pm$\,130  & \pp0.4\,$\pm$\,0.5 &  2.1\,$\pm$\,0.5 & $-$0.54\,$\pm$\,0.41 \\



  \hline
  \end{tabular}
  \end{center}
\end{table*}

\subsection{Stellar Parameters and Metallicity} 
\label{sub:stellar_parameters_and_metallicity}

Table~\ref{tb:stellar-params} summarises the derived stellar parameters.
For the remainder of this paper, when discussing stellar parameters,
we adopt those derived using the 24-arm telluric method
(i.e. the results in the left-hand part of Table~\ref{tb:stellar-params}).
The average metallicity for our sample of 11 RSGs in NGC\,6822 is
[$\bar{\rm Z}$]~=~$-$0.52\,$\pm$\,0.21.
This result is in good agreement with the average metallicity derived in
NGC\,6822 from blue supergiant stars
\citep[BSGs;][]{1999A&A...352L..40M,2001ApJ...547..765V}.

A direct comparison with metallicities from BSGs is legitimate as the results derived here yield a global metallicity ([Z]) which
closely resembles the Fe/H ratio derived by
\cite{2001ApJ...547..765V}.
While our [Z] measurements are also affected by Si and Ti,
we assume [Z]~=~[Fe/H] for the purposes of our discussion.
Likewise, we can compare oxygen abundances (relative to solar) obtained from HII regions as a proxy for [Z] by
introducing the solar oxygen abundance
{12\,+\,log\,(O/H)}$_{\odot}$~=~8.69
\citep{2009ARA&A..47..481A} through the relation
[Z]~=~12\,+\,log\,(O/H)\,$-$\,8.69.

The RSG and BSG stages are different evolutionary phases within the life cycle of a massive star,
while HII regions are the birth clouds which give rise to the youngest stellar population.
As the lifetimes of RSGs and BSGs are $<50$\,Myr,
their metallicity estimates are also expected to be representative of their birth clouds.

\begin{figure}
\includegraphics[width=9.0cm]{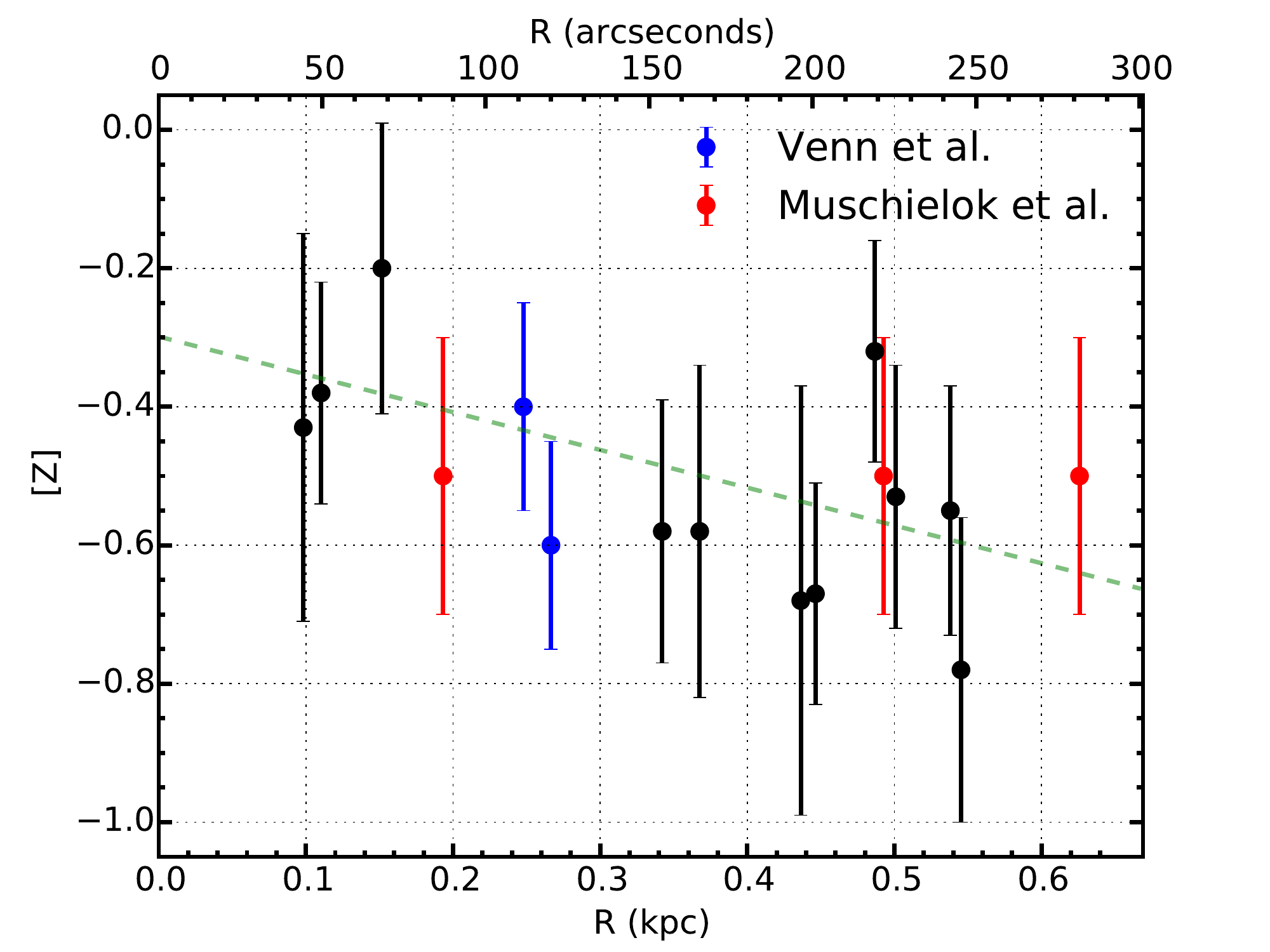}
\caption{
Derived metallicities for 11 RSGs in NGC\,6822 shown against their distance from the galaxy centre;
the average metallicity is
[$\bar{\rm Z}$]~=~$-$0.52\,$\pm$\,0.21.
Blue and red points show BSG results from
\protect\cite{2001ApJ...547..765V} and
\protect\cite{1999A&A...352L..40M}, respectively.
A least-squares fit to the KMOS results reveals a low-significance abundance gradient
(see text for details).
For comparison,
R$_{25}$~=~460$\arcsec$
\citep[=~1.03\,kpc;][]{2012AJ....144....4M}.
        }
\label{fig:ZvsR}
\end{figure}

To investigate the spatial distribution of chemical abundances in NGC\,6822,
in Figure~\ref{fig:ZvsR}
we show the metallicities of our RSGs as a function of radial distance from the centre of the galaxy,
as well as the results from
\cite{2001ApJ...547..765V} and the indicative estimates from
\cite{1999A&A...352L..40M}.

A least-squares fit to the KMOS results reveals a low-significance abundance gradient within the central 1\,kpc of NGC\,6822 of $-$0.5\,$\pm$\,0.4\,dex\,kpc$^{-1}$.
The extrapolated central metallicity from the fit (i.e. at R~=~0) of [Z]~=~$-$0.30\,$\pm$\,0.15 derived remains consistent with the average metallicity assuming no gradient.

\begin{figure}
\includegraphics[width=9.0cm]{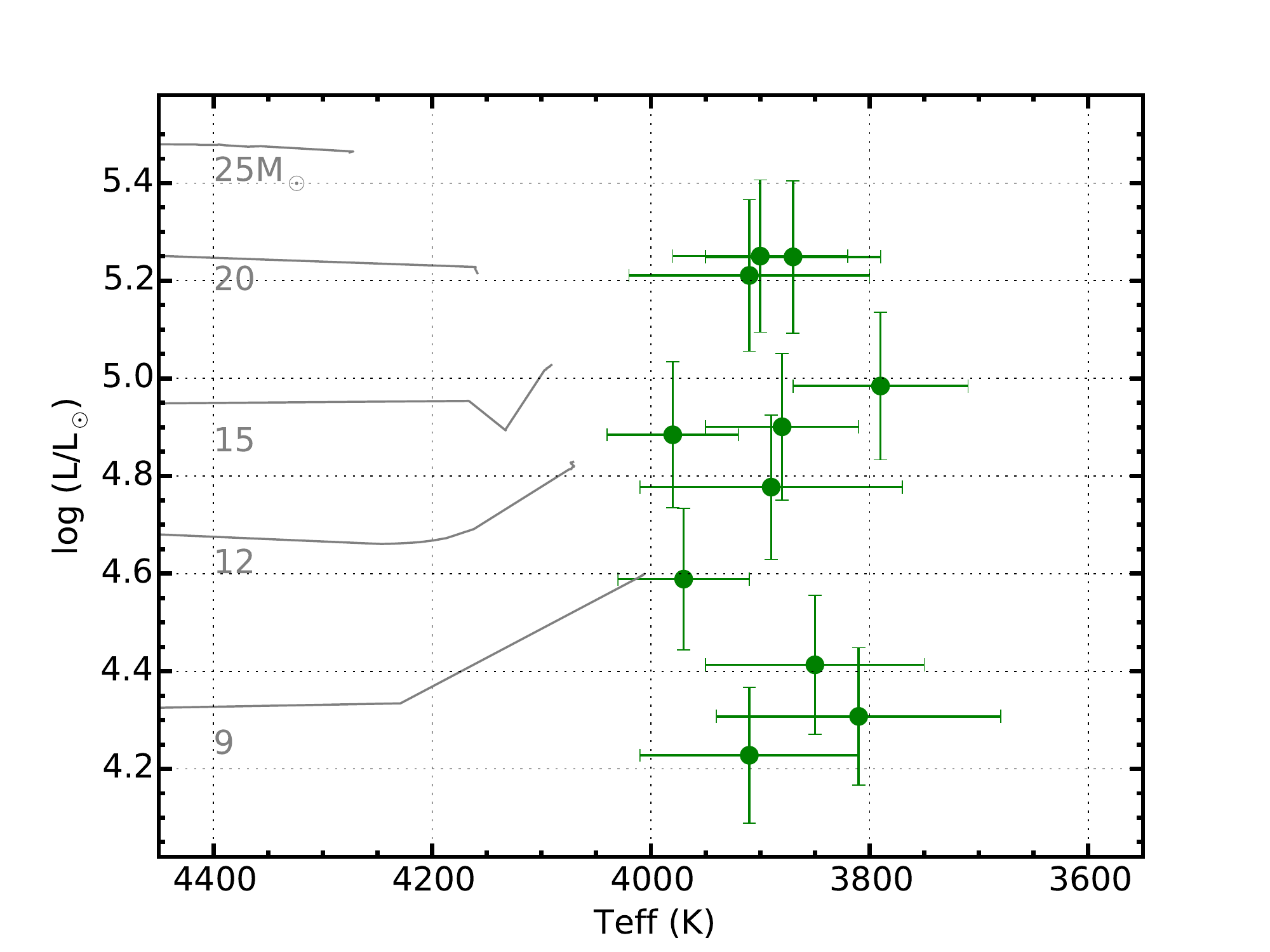}
\caption{
H\textendash R diagram for the 11 RSGs in NGC\,6822.
Evolutionary tracks including rotation
($v/v_{c}$~=~0.4) for SMC-like metallicity (Z~=~0.002)
are shown in grey, along with their zero-age mass
\protect\citep{2013A&A...558A.103G}.
Bolometric corrections are computed using the calibration from
\protect\cite{2013ApJ...767....3D}.
We note that, compared with the evolutionary tracks,
the observed temperatures of NGC\,6822 RSGs are systematically cooler.
This is discussed in Section~\ref{sub:temperatures_of_rsgs}.
}
\label{fig:6822HRD}
\end{figure}

Figure~\ref{fig:6822HRD} shows the location of our sample in the Hertzsprung\textendash Russell
(H\textendash R) diagram.
Bolometric corrections were computed using the calibration from
\cite{2013ApJ...767....3D} to calculate luminosities.
This figure shows that the temperatures derived using the $J$-band method are systematically cooler than the end of the evolutionary models (which terminate at the end of the carbon-burning phase for massive stars) for
Z~=~0.002 from
\cite{2013A&A...558A.103G}.
This is discussed in Section~\ref{sub:temperatures_of_rsgs}.

\section{Discussion} 
\label{sec:discussion}

\subsection{Metallicity Measurements} 
\label{sub:metallicity_measurements}

We find an average metallicity for our sample of
[$\bar{\rm Z}$]~=~$-$0.52\,$\pm$\,0.21
which agrees well with the results derived from BSGs
\citep{1999A&A...352L..40M,2001ApJ...547..765V,Przybilla02} and HII regions
\citep{2006ApJ...642..813L}.

We also find evidence for a low-significance metallicity gradient within the central 1\,kpc of NGC\,6822
($-$0.5\,$\pm$\,0.4\,dex\,kpc$^{-1}$ with a
$\chi^{2}_{red}$~=~1.16; see Figure~\ref{fig:ZvsR}).
The gradient derived is consistent with the trend reported in
\cite{2001ApJ...547..765V}
from their results for the two BSGs compared with HII regions from
\cite{1980MNRAS.193..219P} and two planetary nebulae from
\cite{1995ApJ...445..642R} at larger galactocentric distances.
Our result is also consistent with the gradient derived from a sample of 49 local star-forming galaxies
(Ho et al. submitted).
Including the results for BSGs from
\cite{2001ApJ...547..765V}
in our analysis
gives a consistent gradient
($-$0.48\,$\pm$\,0.33\,dex\,kpc$^{-1}$)
with a
$\chi^{2}_{red}$~=~1.06.
Results from
\cite{1999A&A...352L..40M} are not included in the fit as these measurements were qualitative estimates of metallicity.

In contrast,
\cite{2006ApJ...642..813L} used the oxygen abundances from 19 HII
regions and found no clear evidence for a metallicity gradient.
Using a subset of the highest quality HII
region data available, these authors found a gradient of
$-$0.16\,$\pm$\,0.05\,dex\,kpc$^{-1}$.
Including these results into our analysis degrades the fit and changes the derived gradient significantly
($-$0.18\,$\pm$\,0.05\,dex\,kpc$^{-1}$;
$\chi^{2}_{red}$~=~1.78).
At this point it is not clear whether the indication of a gradient obtained from the RSGs and BSGs is an artefact of the small sample size,
or indicates a difference with respect to the HII region study.

There have been a number of studies of the metallicity of the older stellar population in NGC\,6822.
From spectroscopy of red giant branch (RGB) stars,
\cite{2001MNRAS.327..918T} found a mean metallicity of [Fe/H]~=~$-$0.9
with a reasonably large spread (see their Fig. 19).
More recently,
\cite{2012A&A...540A.135S} derived metallicities using a population of AGB stars within the central 4\,kpc of NGC\,6822.
They found an average metallicity of [Fe/H]~=~$-$1.29\,$\pm$\,0.07\,dex.
Likewise,
\cite{2013ApJ...779..102K}
used spectra of red giant stars within the central 2\,kpc and found an average metallicity of
[Fe/H]~=~$-$1.05\,$\pm$\,0.49\,dex.
We note that none of these studies found any compelling evidence for spatial variations in the stellar metallicities,
which is not surprising given that, in disc galaxies, radial migration is thought to smooth out any abundance gradients over time.
The stellar populations used for these studies are known to be substantially older than our sample;
therefore, owing to the chemical evolution in the time since the birth of the older populations,
we expect the measured metallicities to be significantly lower.

The low metallicity of the young stellar population and the interstellar medium (ISM) in NGC\,6822 can be understood as a consequence of the fact that it is a relatively gas-rich galaxy with a mass of
M$_{HI}$~=~1.45\,$\times$\,10$^{8}$\,M$_{\odot}$
\citep{2004AJ....128...16K} and a total stellar mass of
M$_{*}$~=~0.83\,$-$\,1.70\,$\times$\,10$^{8}$\,M$_{\odot}$
\citep{2008MNRAS.390.1453W,2013ApJ...779..102K,2014ApJ...789..147W}.

The simple closed-box chemical-evolution model relates the metallicity mass fraction $Z(t)$ at any time to the ratio of stellar to gas mass $M_{*}\over M_{g}$ through:

\begin{equation}\label{closed-box}
Z(t) = {y \over 1-R } \ln \left[ 1 + {M_{*}(t)\over M_{g}(t)}  \right],
\end{equation}

\noindent where $y$ is the fraction of metals per stellar mass produced through stellar nucleosynthesis
(the so-called yield) and $R$ is the fraction of stellar mass returned to the ISM through stellar mass-loss.

According to
Kudritzki et al. (in prep.), the ratio $y/(1-R)$ can be empirically determined from the fact that the metallicity of the young stellar population in the solar neighbourhood is solar, with a mass fraction of Z$_{\odot}$~=~0.014
\citep{2012A&A...539A.143N}.
With a stellar-to-gas mass column density of 4.48 in the solar neighbourhood
\citep{2003ApJ...587..278W,2013ApJ...779..115B}
one then obtains $y/(1-R)$~=~0.0082~=~0.59Z$_{\odot}$ with an uncertainty of 15\% dominated by the 0.05\,dex uncertainty of the metallicity determination of the young population.

Accounting for the presence of helium and metals in the neutral interstellar gas we can turn the observed HI mass in NGC\,6822 into a gas mass via
M$_{g}$~=~1.36\,M$_{HI}$ and use the simple closed-box model to predict a metallicity of
[Z]~=~$-$0.44 to $-$0.69,
in good agreement with our value obtained from RSG spectroscopy.

As discussed above, the older stellar population of AGB stars has a metallicity roughly 0.8\,dex lower than the RSGs.
In the framework of the simple closed-box model this would correspond to a period in time where the ratio of stellar to gas mass was $\sim$0.1
(much lower than the present value of 0.42 to 0.86).
The present star-formation rate of NGC\,6822 is $\sim$0.02 M$_{\odot}$yr$^{-1}$
\citep{2010A&A...512A..68G,2011ApJ...730...88E}.
At such a level of star formation it would have taken 5\,Gyr to produce the presently observed stellar mass and to arrive at the average metallicity of the young stellar population from that of the AGB stars
(of course, again relying on the simple closed-box model).
Evidence suggests that the star-formation rate was substantially lower in the past
\citep{2011ApJ...730...88E,2014ApJ...789..147W},
therefore, the build up of the observed stellar mass would have taken correspondingly longer.

Given the irregularities present in the stellar and gaseous morphology of NGC\,6822,
this galaxy may not be a good example of a closed-box system;
however, it is remarkable that the closed-box model reproduces the observed metallicity so closely.


\subsection{Temperatures of RSGs} 
\label{sub:temperatures_of_rsgs}

Effective temperatures have been derived for 11 RSGs from our observed sample in NGC\,6822.
To date, this represents the fourth data set used to derive stellar parameters in this way and the first with KMOS.
The previous three data sets which have been analysed are those of 11 RSGs in Perseus OB-1,
a Galactic star cluster
\citep{2014ApJ...788...58G}, nine RSGs in the LMC and 10 RSGs in the SMC
\citep[both from][]{Davies-prep}.
These results range from Z~=~Z$_{\odot}$ in Perseus OB-1 to Z~=~0.3\,Z$_{\odot}$ in the SMC, around 0.5\,dex in metallicity.

\begin{figure}
\includegraphics[width=9.0cm]{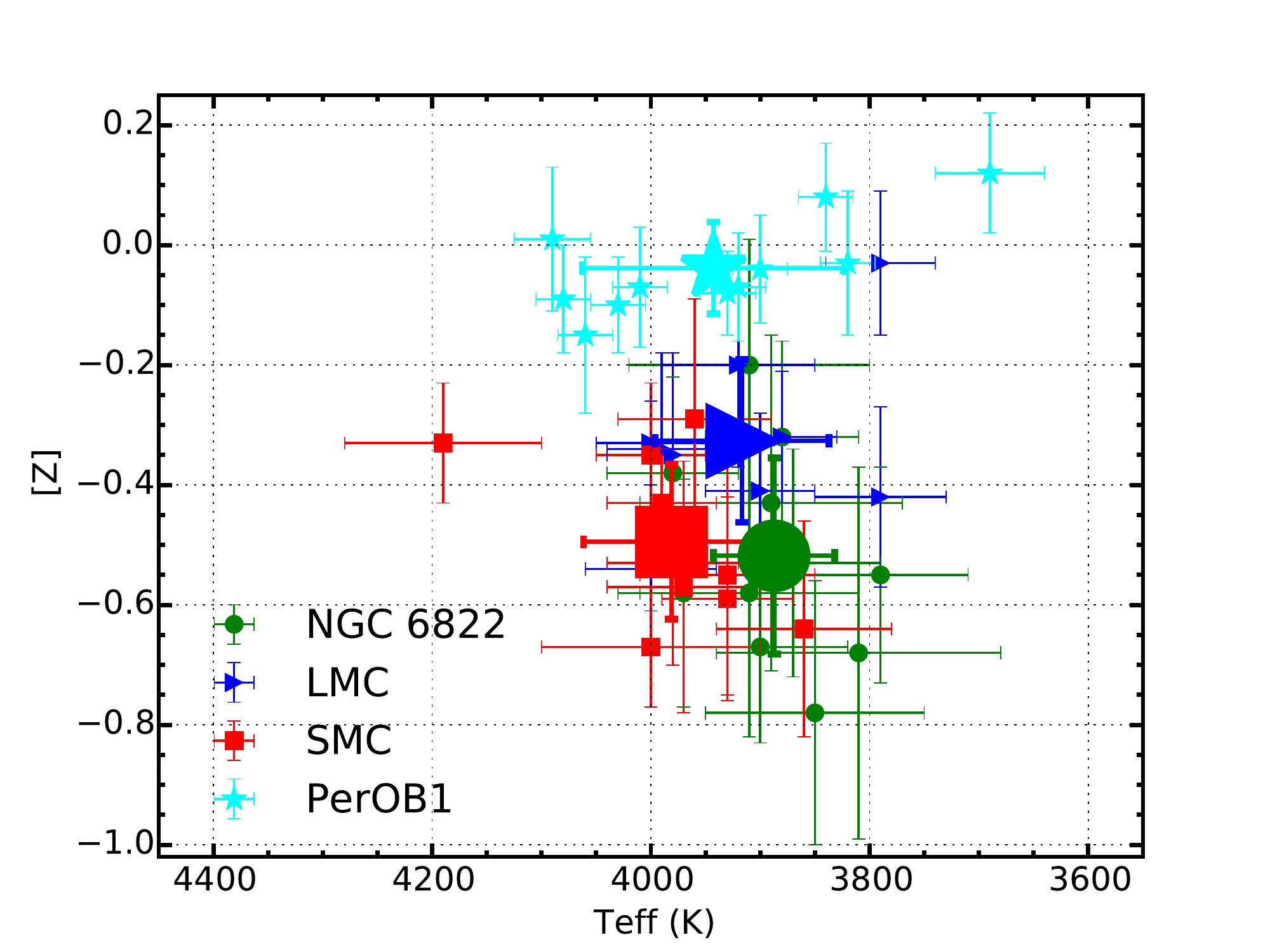}
\caption{
Effective temperatures as a function of metallicity for four different data sets using the $J$-band analysis technique.
There appears to be no significant variation in the temperatures of RSGs over a range of 0.55\,dex.
These results are compiled from the LMC, SMC
\protect\citep[blue and red points, respectively;][]{Davies-prep}, Perseus OB-1
\protect\citep[a Galactic RSG cluster; cyan;][]{2014ApJ...788...58G} and NGC\,6822 (green).
Mean values for each data set are shown as enlarged points in the same style and colour.
The x-axis is reversed for comparison with Figure~\ref{fig:HRD}.\label{fig:TvsZ}
\vspace{0.2cm}
  }
\end{figure}

\begin{figure}
\includegraphics[width=9.0cm]{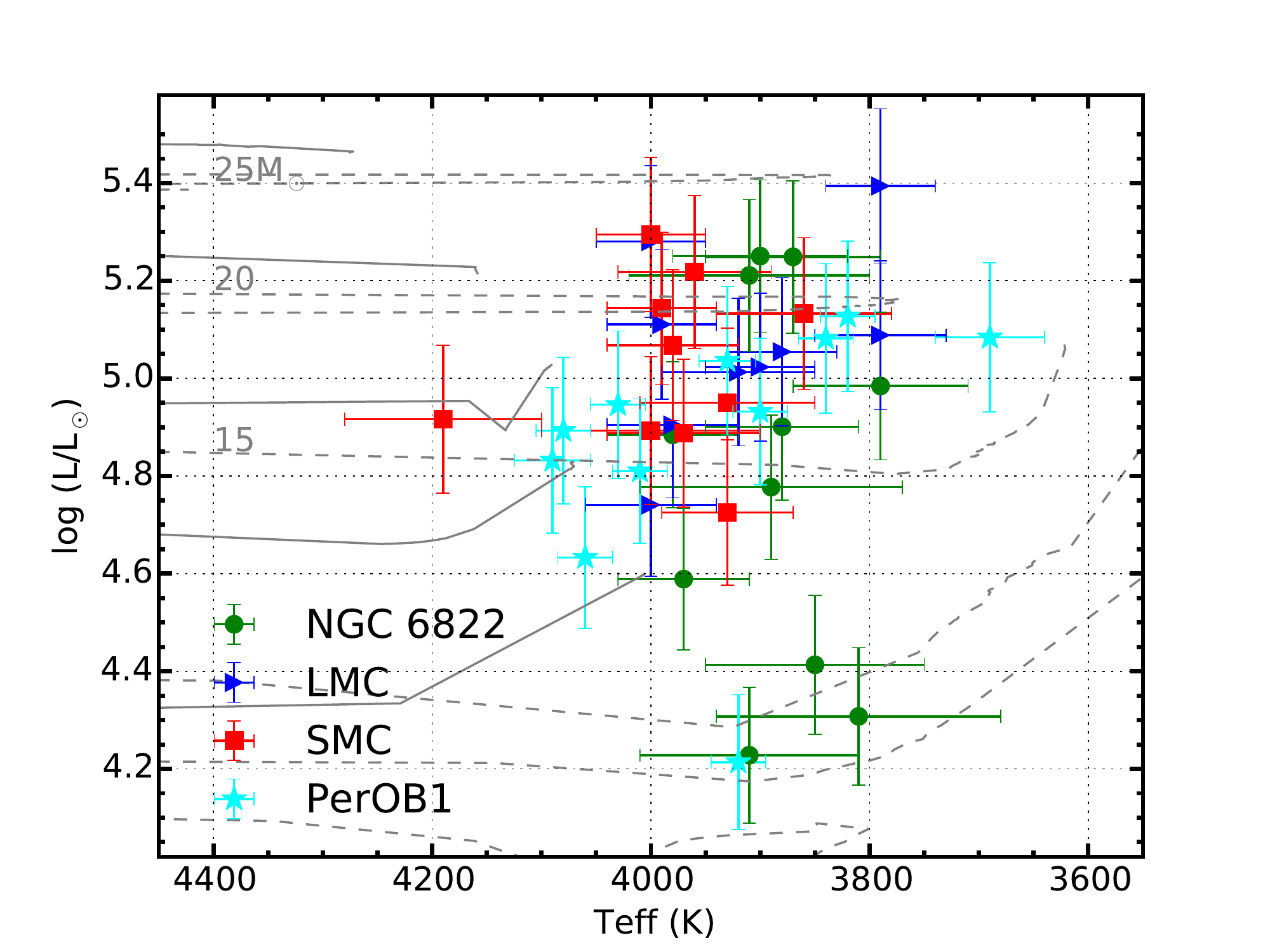}
\caption{
H\textendash R diagram for RSGs in Perseus OB-1 (cyan), LMC (blue), SMC (red) and NGC\,6822 (green) which have stellar parameters obtained using the $J$-band method.
This figure shows that there appears to be no significant temperature differences between the four samples.
Solid grey lines show SMC-like metallicity evolutionary models including rotation
\protect\citep{2013A&A...558A.103G}.
Dashed grey lines show solar metallicity evolutionary models including rotation
\protect\citep{2012A&A...537A.146E}.\label{fig:HRD}
\vspace{0.5cm}
        }
\end{figure}

We compare the effective temperatures derived in this study with those of the previous results in different environments.
Their distribution is shown as a function of metallicity in Figure~\ref{fig:TvsZ}.
Additionally, Figure~\ref{fig:HRD} shows the H\textendash R diagram for the four sets of results.
Bolometric corrections to calculate the luminosities for each sample were computed using the calibration from
\cite{2013ApJ...767....3D}.

From these figures, we see no evidence for significant variations in the average temperatures of RSGs with respect to metallicity.
This is in contrast with current evolutionary models which display a change of $\sim$450\,K,
for a M~=~15M$_{\odot}$ model,
over a range of solar to SMC-like metallicities~\citep{2012A&A...537A.146E,2013A&A...558A.103G}.

For solar metallicity, observations in Perseus OB-1 are in good agreement with the models
\citep[see Fig. 9 in][]{2014ApJ...788...58G}.
However, at SMC-like metallicity, the end-points of the models are systematically warmer than the observations.
The temperature of the end-points of the evolutionary models of massive stars could depend on the choice of the convective mixing-length parameter
\citep{1992A&AS...96..269S}.
The fact that the models produce a higher temperature than observed could imply that the choice of a solar-like mixing-length parameter does not hold for higher mass stars at lower metallicity.

Lastly, we note that the average spectral type of RSGs tends towards an earlier spectral type with decreasing metallicity over this range
\citep{1979ApJ...231..384H,2012AJ....144....2L}.
We argue that this is not in contradiction to the above results.
Spectral types are derived for RSGs using the optical TiO band-heads at
$\sim$0.65\,$\mu$m,
whereas in this study temperatures are derived using near-IR atomic features
(as well as the line-free pseudo-continuum).
The strengths of TiO bands are dependent upon metallicity which means that
the spectral classification for RSGs at a constant temperature will differ
\citep{2013ApJ...767....3D}.
Therefore, although historically spectral type has been used as a proxy for temperature, this assumption does not provide an accurate picture for RSGs in differing environments.

\subsection{AGB Contamination} 
\label{sub:AGB_contamination}
As mentioned in Section~\ref{sub:target_selection}, massive AGB stars are potential contaminants to our sample.
These stars have similar properties to RSGs and can occupy similar mass ranges as lower-mass RSGs
\citep{2005ARA&A..43..435H};
however, their lifetimes are around $\sim$250\,Myr
\citep{2010MNRAS.401.1453D}.
\cite{1983ApJ...272...99W} argued for an AGB luminosity limit
(owing to the limit on the mass of the degenerate core) of M$_{bol}\sim-$7.1.
Using this maximum luminosity, corrected for the distance to NGC\,6822,
yields $K$~=~14.0.
Four of our analysed stars have $K$-band magnitudes fainter than this limit,
but excluding the results for these does not significantly alter any of our results.
However, given the near-IR luminosity and colours of these stars, if indeed they are AGB stars, they are probably younger massive AGB stars
\citep[e.g.][]{2008IAUS..252..297S} and therefore would arguably still trace the relatively young stellar population.

\section{Conclusions} 
\label{sec:conclusions}

KMOS spectroscopy of RSG stars in NGC\,6822 is presented.
The data were telluric corrected in two different ways and the standard three-arm telluric method is shown to work as effectively (in most cases) as the more time expensive 24-arm telluric method.
Radial velocities of the targets are derived and are shown to be consistent with previous results in NGC\,6822, confirming their extragalactic nature.

Stellar parameters are calculated for 11 RSGs using the $J$-band analysis method outlined by
\cite{2010MNRAS.407.1203D}.
The average metallicity within NGC\,6822 is
[$\bar{\rm Z}$]~=~$-$0.52\,$\pm$\,0.21,
consistent with previous abundance studies of young stars.
We find an indication of a metallicity gradient within the central 1\,kpc,
but with a low significance caused by the small size and limited spatial extent of our RSG sample.
To conclusively assess the presence of a metallicity gradient among the young population within NGC\,6822 a larger systematic study of RSGs is needed.

The chemical abundances of the young and old stellar populations of NGC\,6822 are well explained by a simple closed-box chemical evolution model.
However, while an interesting result, we note that the closed-box model is unlikely to be a good assumption for this galaxy given its morphology.

The effective temperatures of those in this study are compared with RSGs analysed using the same method in previous studies.
Using results which span 0.55\,dex in metallicity (solar to SMC) within four galaxies, we find no evidence for a systematic variation in average effective temperature with respect to metallicity.
This is in contrast with evolutionary models which, for a  similar change in metallicity,
predict a shift in the temperature of RSGs of up to 450\,K.
We argue that an observed shift in average spectral type of RSGs over this metallicity range does not imply a shift in average temperature.

These observations were taken as part of the KMOS Science Verification programme.
With guaranteed time observations we have obtained data for RSGs in NGC\,300 and NGC\,55 at distances of $\sim$1.9\,Mpc,
as well as observations of super-star clusters in M\,83 and the Antennae galaxy at 4.5 and 20\,Mpc, respectively.
Owing to the fact that RSGs dominate the light output from super-star clusters
\citep{2013MNRAS.430L..35G} these clusters can be analysed in a similar manner
\citep{2014ApJ...787..142G},
which will provide metallicity measurements at distances a factor of 10 larger than using individual RSGs!
This work is the first step towards an ambitious proposal to survey a large number of galaxies in the Local Volume,
motivated by the twin goals of investigating their abundance patterns,
while also calibrating the relationship between galaxy mass and metallicity in the Local Group.


\acknowledgments

We thank the referee for a thorough review and constructive comments.
We thank Mike Irwin for providing the photometric catalogue from the WFCAM observations.
RPK and JZG acknowledge support by the National Science Foundation under grant AST-1108906.
The NSO/Kitt Peak FTS high-resolution telluric spectra used here to
investigate the wavelength solution were produced by NSF/NOAO.

{\it Facilities:}
\facility{VLT (KMOS)}.



\end{document}